\documentclass[12pt,preprint]{aastex}
\usepackage{natbib}
\usepackage{ifthen}
\newcounter{address}
\newcommand{\latin}[1]{\textit{#1}}
\newcommand{\ie}{\latin{ie}}

\newcommand{\etal}{\latin{et~al}}
\newlength{\threewidth}
\newlength{\twowidth}
\newlength{\twothreewidth}
\newlength{\onewidth}
\newcommand{\Nside}{N_{\mathrm{side}}}
\newcommand{\unit}[1]{\mathrm{#1}}
\renewcommand{\mag}{\unit{mag}}
\newcommand{\rad}{\unit{rad}}
\renewcommand{\arcsec}{\unit{arcsec}}
\newcommand{\ster}{\unit{ster}}
\newcommand{\percent}{\unit{percent}}
\newcommand{\Tycho}{Tycho-2}
\newcommand{\USNOB}{USNO-B Catalog}
\newcommand{\TWOMASS}{2MASS PSC Catalog}
\newcommand{\merged}{merged \Tycho\ and \USNOB s}
\newcommand{\an}{\textsl{Astrometry.net}}

\newcommand{\numSpikes}{24,148,382}
\newcommand{\numHalos}{196,133}
\newcommand{\percentSpikes}{$2.3\,\percent$}  % 2.316128818
\newcommand{\percentHalos}{$0.02\,\percent$}  % 0.01876555

\begin{document}
\title{
  Cleaning the USNO-B Catalog
  through automatic detection of optical artifacts
}
\author{
  Jonathan~T.~Barron\altaffilmark{\ref{Toronto}},
  Christopher~Stumm\altaffilmark{\ref{Toronto}},
  David~W.~Hogg\altaffilmark{\ref{NYU},\ref{email}},
  Dustin~Lang\altaffilmark{\ref{Toronto}},
  Sam~Roweis\altaffilmark{\ref{Toronto},\ref{Google}}
}

\setcounter{address}{1}
\altaffiltext{\theaddress}{\stepcounter{address}\label{Toronto}
Department of Computer Science, University of Toronto,
6 King's College Road, Toronto, Ontario, M5S~3G4 Canada}
\altaffiltext{\theaddress}{\stepcounter{address}\label{NYU}
Center for Cosmology and Particle Physics, Department of Physics, New
York University, 4 Washington Place, New York, NY 10003}
\altaffiltext{\theaddress}{\stepcounter{address}\label{email}
To whom correspondence should be addressed: \texttt{david.hogg@nyu.edu}}
\altaffiltext{\theaddress}{\stepcounter{address}\label{Google}
Google, Mountain View, CA}

\begin{abstract}
The USNO-B Catalog contains spurious entries that are caused by
diffraction spikes and circular reflection halos around bright stars
in the original imaging data. These spurious entries appear in the
Catalog as if they were real stars; they are confusing for some
scientific tasks.  The spurious entries can be identified by simple
computer vision techniques because they produce repeatable patterns on
the sky. Some techniques employed here are variants of the Hough
transform, one of which is sensitive to (two-dimensional)
overdensities of faint stars in thin right-angle cross patterns
centered on bright ($<13\,\mag$) stars, and one of which is sensitive
to thin annular overdensities centered on very bright ($<7\,\mag$)
stars.  After enforcing conservative statistical requirements on
spurious-entry identifications, we find that of the 1,042,618,261
entries in the USNO-B Catalog, 24,148,382 of them ($2.3\,\percent$)
are identified as spurious by diffraction-spike criteria and 196,133
($0.02\,\percent$) are identified as spurious by reflection-halo
criteria.  The spurious entries are often detected in more than 2
bands and are not overwhelmingly outliers in any photometric
properties; they therefore cannot be rejected easily on other grounds,
\ie, without the use of computer vision techniques.  We demonstrate our
method, and return to the community in electronic form a table of spurious
entries in the Catalog.
\end{abstract}


\section{Introduction}

The \USNOB\ \citep{monet03a} is an astrometric catalog containing
information on $\sim10^{9}$ stars.  The original imaging data taken
for this catalog come exclusively from photographic plates, taken from
several different surveys operating over many decades.  These plates
were uniformly scanned and automated source detection was performed on
the scans.  From the sources detected in the scans, the Catalog was
constructed in a relatively ``inclusive'' way.  The sources were
required to be compact, and to show detections in more than one band
of the five bands ($O,E,J,F,N$) from which the Catalog was constructed.
However, the original plate images contained many artifacts, defects,
trailed satellites, and large, resolved sources such as nearby
galaxies, nebulae, and star clusters.  Some of the entries in the
\USNOB\ do not correspond to real, independent, astronomical sources
but rather to arbitrary parts of extended sources, or fortuitously
coincident (across bands) data defects or artificial features.  Though
compact galaxies can be used along with stars for astrometric science,
the artificial features recorded as stars are at best useless---and at
worst damaging---to scientific projects undertaken with the \USNOB.

That said, the \USNOB\ is a tremendously important and productive tool
as the largest visual ($BRI$) all-sky catalog for astrometric science
available at the present day. Users of the Catalog benefit from its
careful construction, its connection to the absolute astrometric
reference frame, and the long time baseline of its originating data.

Our group is using the Catalog for the ambitious \an\ project
\citep{lang07a} in which we locate ``blind'' the position,
orientation, and scale of images with little, no, or corrupted
astrometric meta-data.  For the \an\ project, we need the input
astrometric catalog to have as few spurious entries as possible.
Indeed, in our early work, most of the
``false positive'' results from our blind astrometry system involved
spurious alignments of linear defects in submitted images with
lines of spurious entries in the \USNOB\ coming from diffraction
spikes near bright stars.  For this reason, we found it necessary to
``clean'' the Catalog of as many spurious entries as we can identify
by their configurations on the two-dimensional plane of the sky. In
what follows we describe how we identified two large classes of
spurious entries, thereby greatly improving the value of the Catalog
for our needs.

The most analogous prior work in the astronomical literature is a
cleaning of the SuperCOSMOS Sky Survey using sophisticated computer
vision and machine learning techniques \citep{storkey04a}. Our work
is less general because we have specialized our detection algorithms
to the specific morphologies of the features we know to be present in
the \USNOB. This specialization is possible because the vetting
procedure employed in the construction of the \USNOB\ has eliminated
most of the defects (satellite trails, dirt, and scratches) that have
unpredictable morphologies. This specialization has the great
advantage that it permits us to detect image defects composed of small
numbers of catalog entries, which would not be statistically
identifiable if we did not have a strong \emph{a priori} model for their
morphologies.

In what follows, we will treat the \USNOB\ as a collection of catalog
``entries'', which are rows in a (large) table.  Most of these entries
correspond to ``stars'', which are hot balls of hydrogen in space, or
compact galaxies, which are extremely distant collections of stars,
but which will also be referred to as ``stars'' because from the point
of view of astrometric calibration they behave the same as stars.
Catalog entries that do not correspond to stars or individual compact
galaxies are considered by us to be ``spurious''.  We identify
some fraction of the spurious entries in the \USNOB\ by exploiting the
repeatable configurations they show around bright stars.

\section{Spurious catalog entries}

The \USNOB\ was constructed from imaging in five bands ($O$, $E$, $J$,
$F$, $N$) at two broad epochs ($O$, $E$ at first epoch, $J$, $F$, $N$
at second), taken with plate centers on a (fairly) regular grid of the
sky.  The plate imaging comprising the original data for the Catalog
is heterogeneous (in camera or survey origin and in data quality); in
order to guard against spurious entries, the construction of the
Catalog required detection of sources in multiple bands. However, some
spurious catalog entries survived this requirement.

\subsection{Diffraction spikes}

\begin{figure}
	\hbox{
		\hbox{\resizebox{\threewidth}{!}{\includegraphics{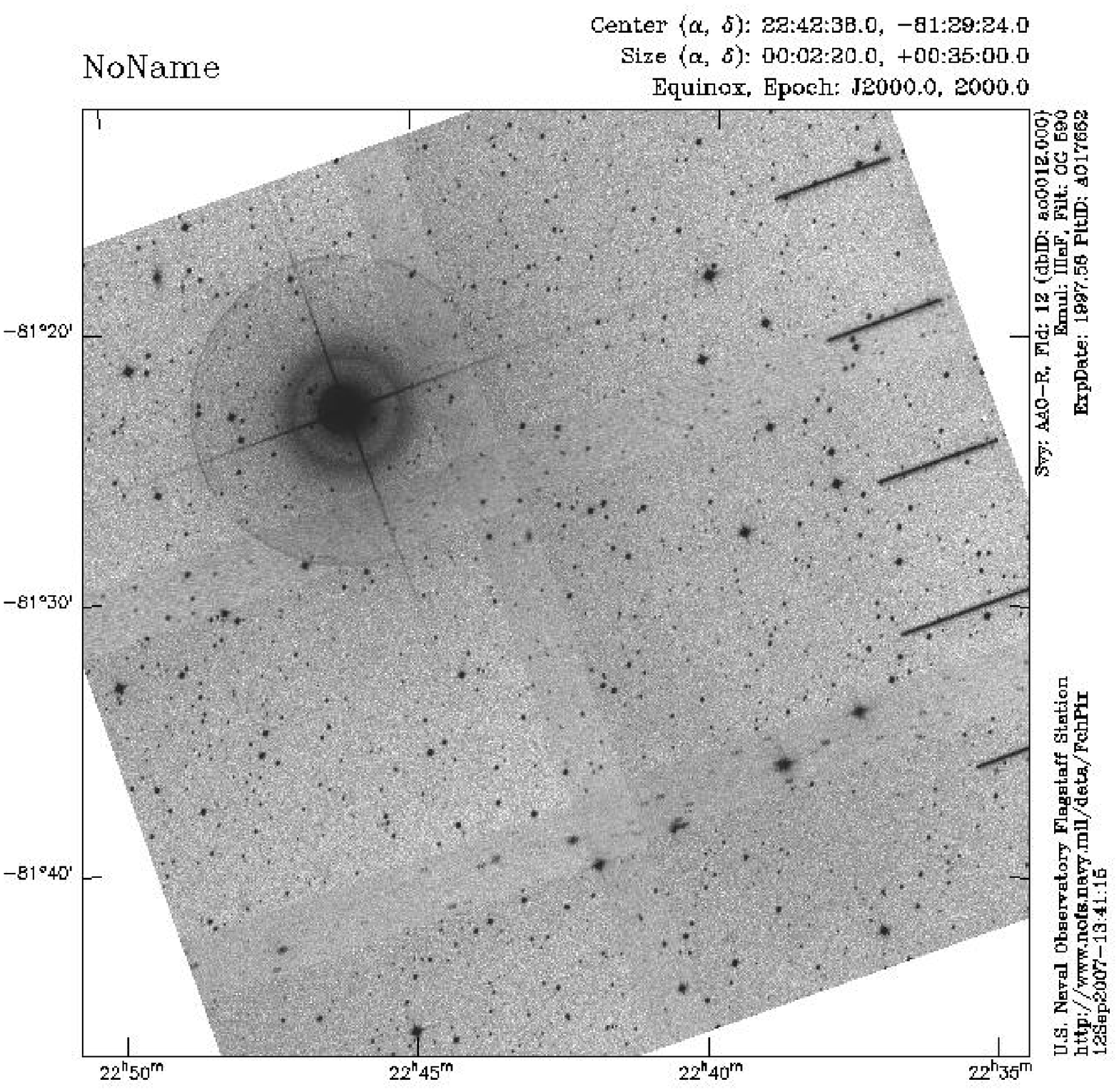}}}
		\hbox{\resizebox{\threewidth}{!}{\includegraphics{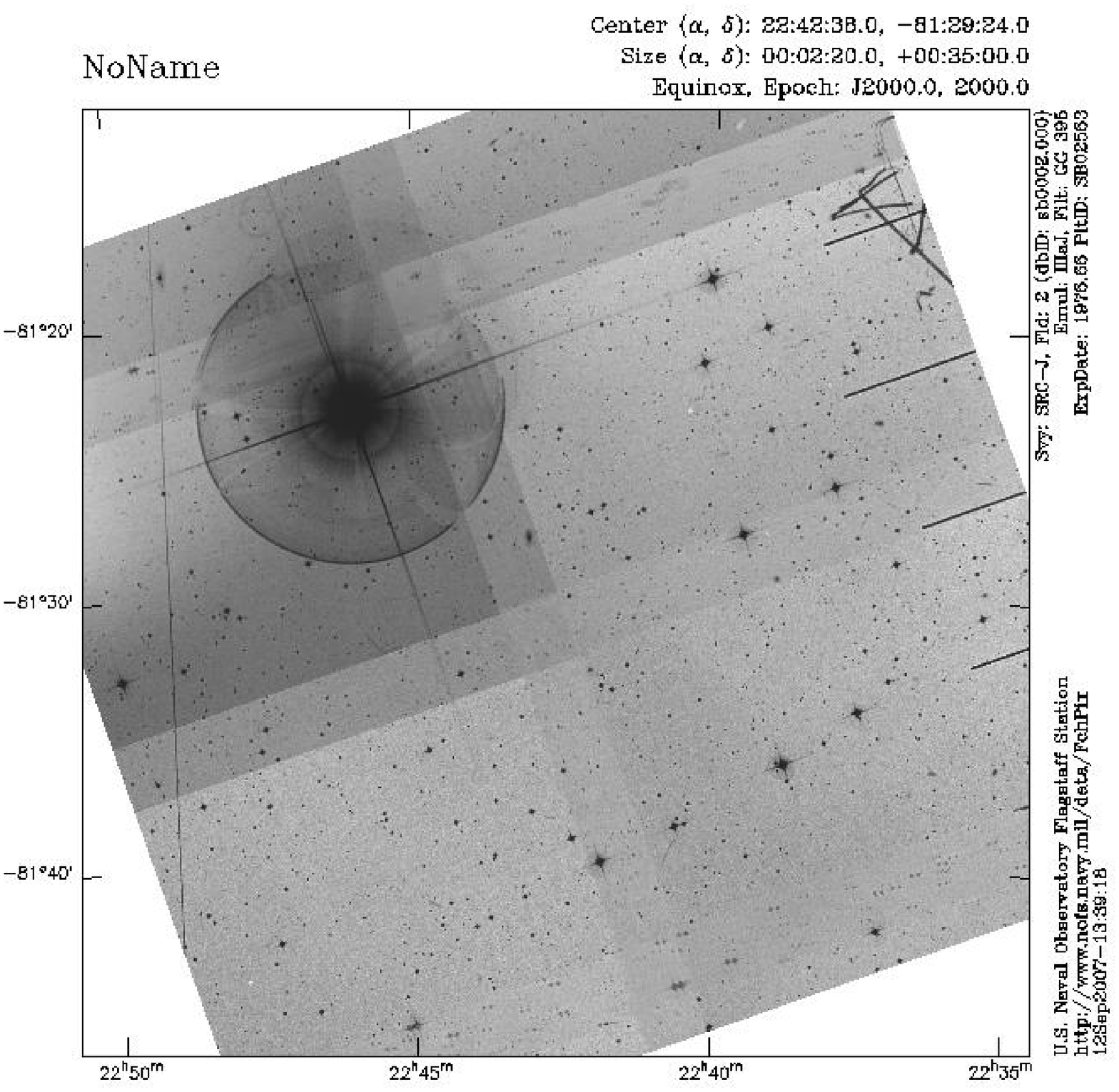}}}
		\hbox{\resizebox{\threewidth}{!}{\includegraphics{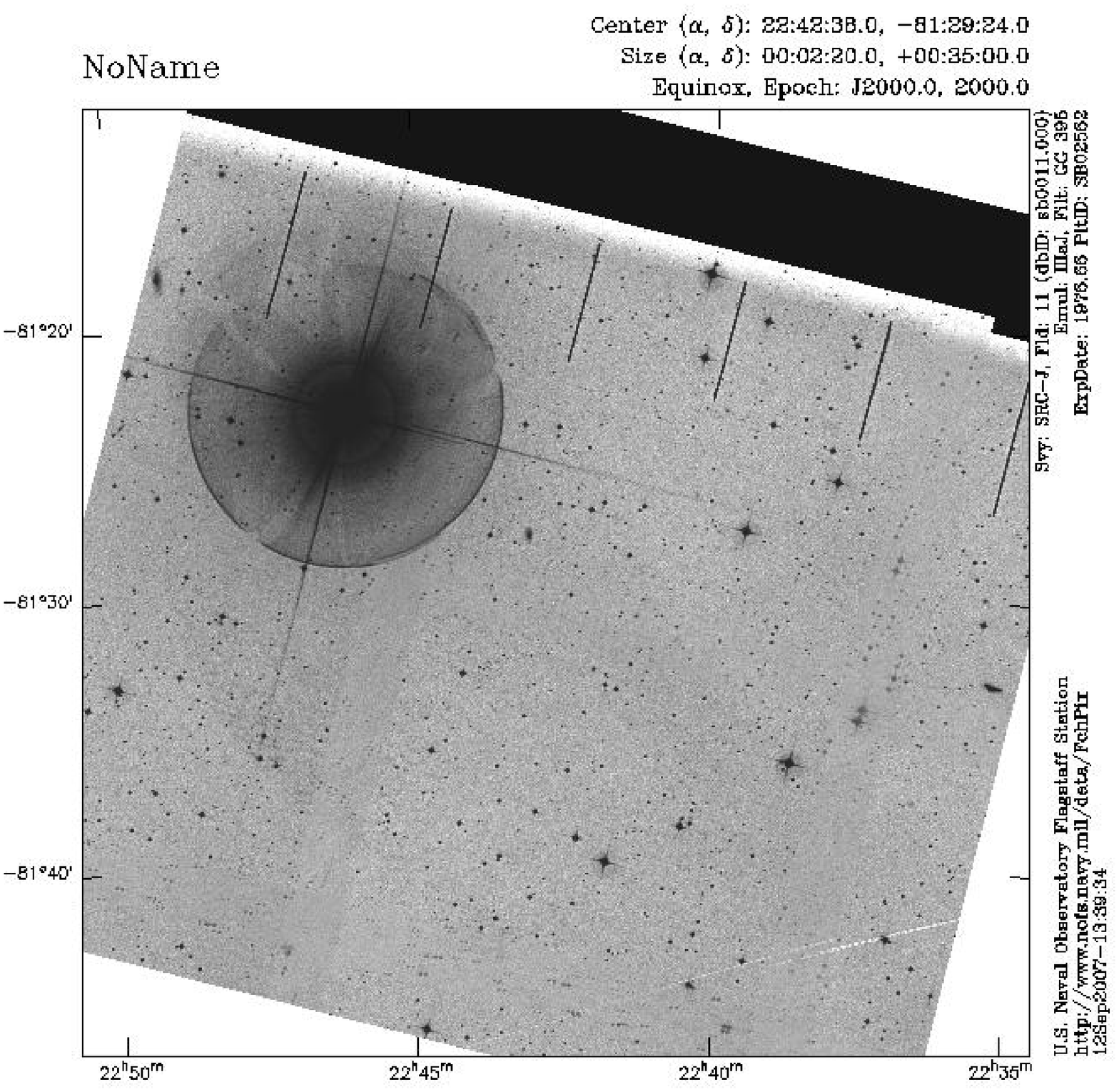}}}
	}
\caption{Subimages of three of the nine scanned plates that overlap a
small patch of sky centered around (RA,Dec)=(341.8, -81.4)~deg (J2000)
from which part of \USNOB was created, retrieved from the US Naval
Observatory Precision Measuring Machine Data Archive. Note the
different orientations of the diffraction spikes generated by brighter
stars, and the multiple halos surrounding the brightest star.
\label{fig:skyPatchSource}}
\end{figure}

\begin{figure}
\hbox{
	\hbox{
		\vbox{
			\hbox{\resizebox{\threewidth}{!}{\includegraphics{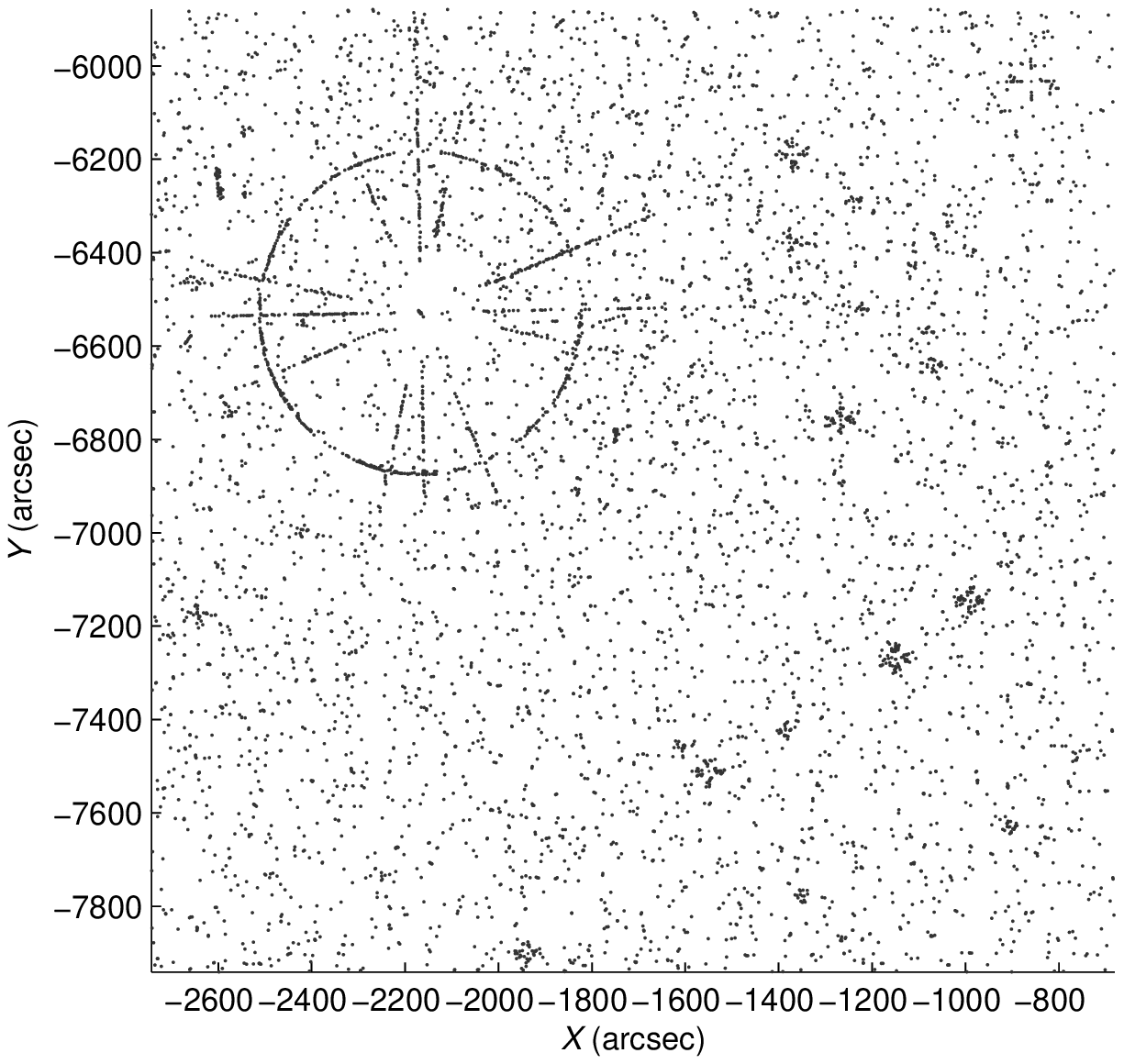}}}
			\hbox{\resizebox{\threewidth}{!}{\includegraphics{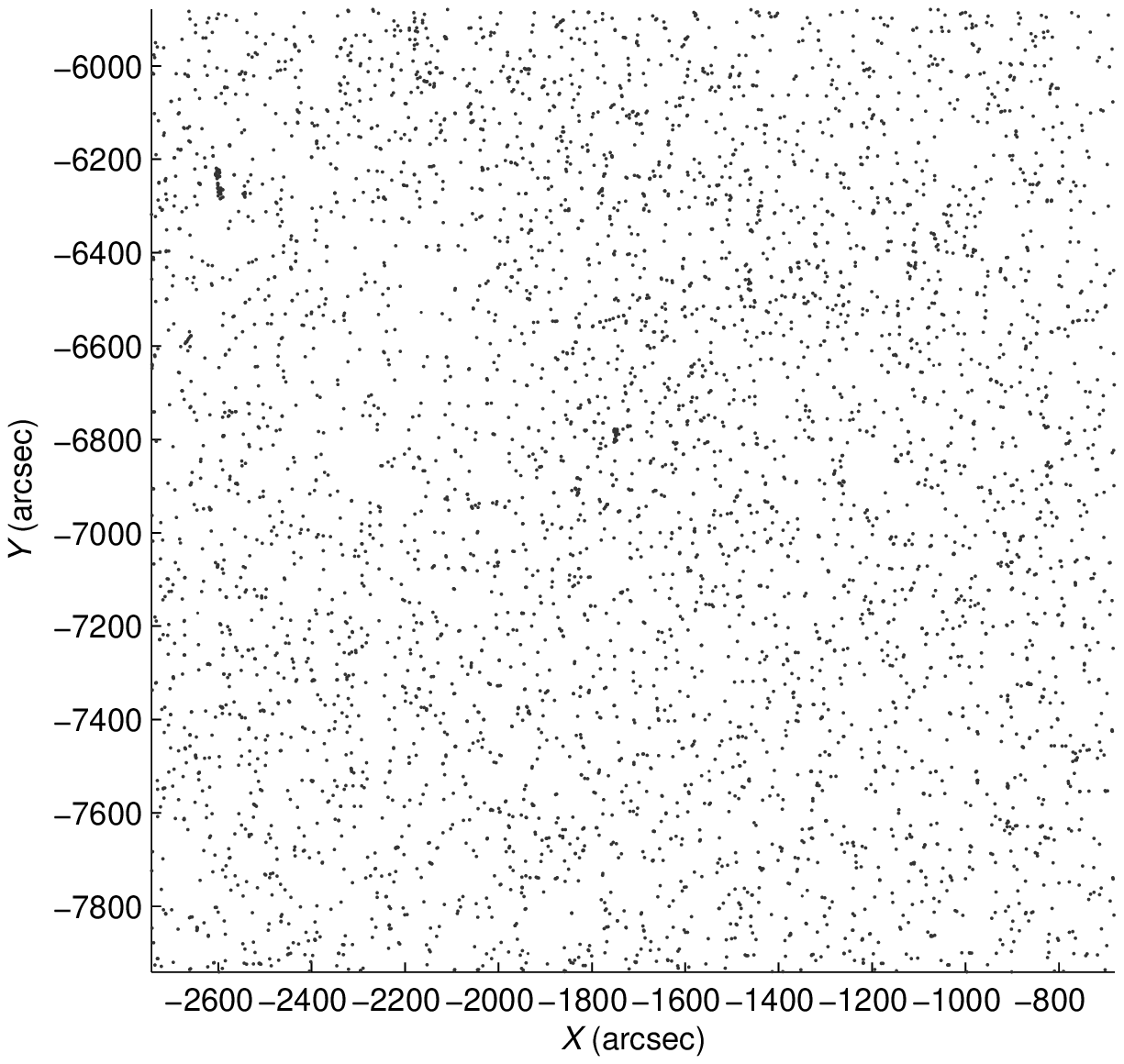}}}
			}
		}
	\hbox{
		\resizebox{\twothreewidth}{!}{\includegraphics{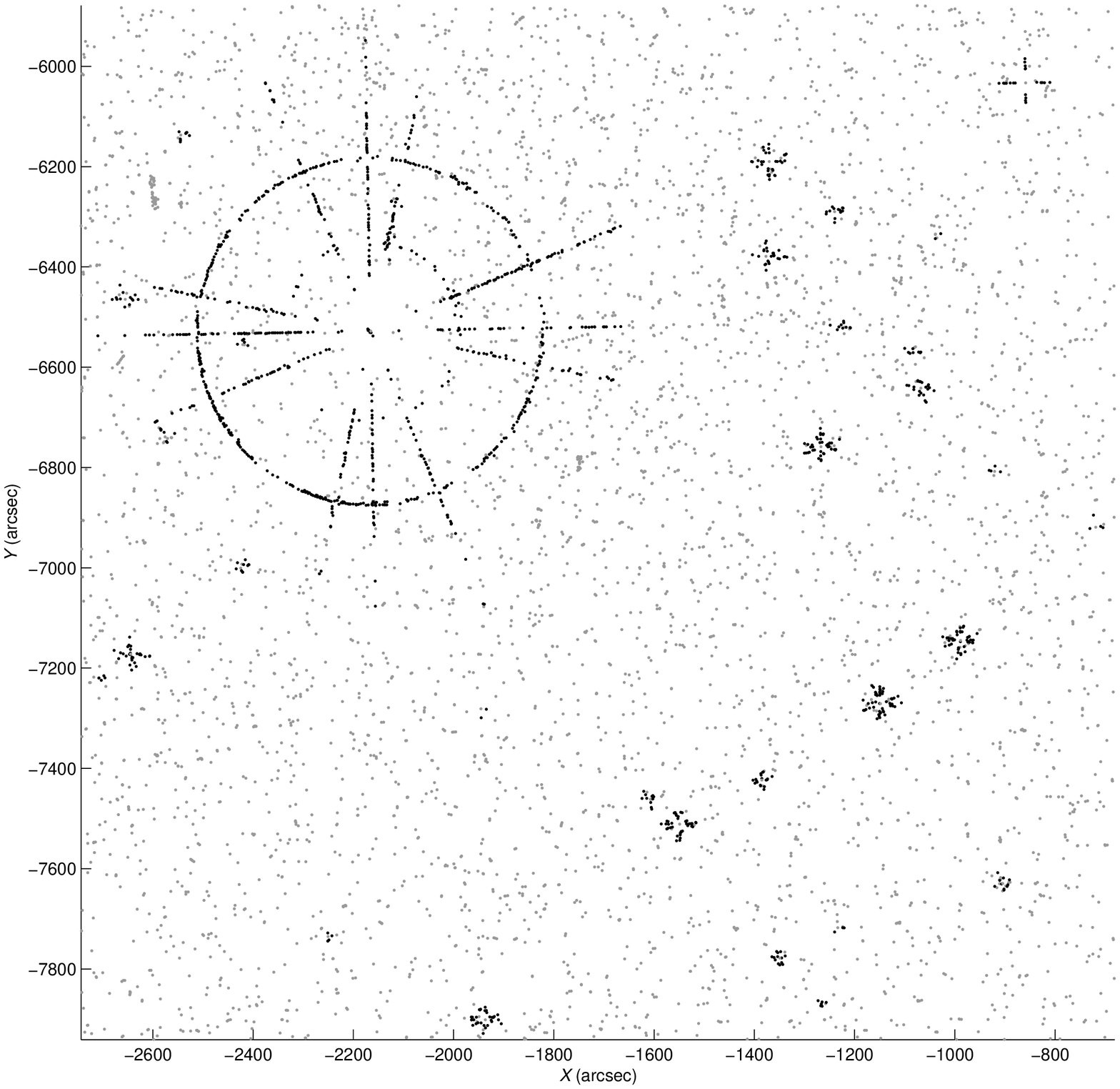}}
	}
}
\caption{The same small patch of sky as pictured in
Figure~\ref{fig:skyPatch}, taken from the \USNOB, in tangent-plane
coordinates relative to a tangent point at the center of the
containing healpixel, in units of $\arcsec$.  The bright stars in this
patch have multiple sets of diffraction spikes because they lie in a
sky region where plates taken at different orientations overlap.
\textsl{Upper left panel:} All \USNOB\ entries in this patch.
\textsl{Right panel:} The same patch with dark points showing catalog
entries marked as spurious by either the diffraction spike or
reflection halo criteria described in the text.  \textsl{Lower left
panel:} The same patch, with only non-spurious entries shown.
\label{fig:skyPatch}}
\end{figure}

The diffraction-limited point-spread function of a physical telescope
is related to the Fourier transform of the entrance aperture. In this
transform, the thin cross-like support structure holding the secondary
mirror in the entrance aperture produces a large cross-like pattern in
the stellar point-spread function (PSF). The sources automatically
extracted from the scans of the photographic plate images include many
spurious features that are in fact just detections of these diffraction
spikes (Figures~\ref{fig:skyPatchSource} and \ref{fig:skyPatch}).

The survey cameras that took the imaging data used to construct the
\USNOB\ are on equatorial mounts and have no capability for rotation
of the support structure relative to the sky once the pointing of the
telescope is set.  The diffraction spikes for any two images 
taken by the same camera at the same pointing are therefore always aligned. For
this reason, spurious stars detected as part of one of these spikes in
one image in one band often line up with spurious stars detected in
the corresponding spike in some other band. Some spurious ``spike''
catalog entries thereby satisfy the \USNOB\ vetting requirement that
catalog entries have cospatial counterparts in multiple bands.

Fortunately, spurious spike entries can be identified on the basis of
morphological regularities in the two-dimensional distribution on the
sky of the spurious catalog entries they generate. These regularities
include the following: \textsl{(1)}~Diffraction spikes are centered on
bright ($<13\,\mag$) stars. In what follows, the central star for a
diffraction spike will be referred to as the ``generating star''.
\textsl{(2)}~Because telescope supports are usually four perpendicular
rods, each diffraction spike generated by a bright star has four lines
at right angles to one another.  \textsl{(3)}~The diffraction spike
brightness is proportional to the brightness of the generating star,
but each spike becomes fainter with angular distance from the
generating star. Given that sources extracted from the scanned plates
are detected to some limiting brightness, the angular length of a
diffraction spike is closely related to the magnitude of the
generating star (Figure~\ref{fig:spikeProperties}).  \textsl{(4)}~The
angular width of a diffraction spike is narrow, so the two-dimensional
density of spurious spike entries can be very large. The angular width
is set by physical optics and is therefore roughly independent of the
magnitude of the generating star.  \textsl{(5)}~The orientation of the
diffraction spike pattern is roughly common to all spikes taken by the
same camera at the same pointing.
We can use the regularities among diffraction spikes to guide a
sensitive, automated search.

Each \USNOB\ entry is tagged with a survey identifier and one or more
field numbers corresponding to the plates in that survey in which it
was detected.  Because all diffraction spikes in one field will share
the same orientation and properties, we analyze the \USNOB\ entries
one field at a time.  In this context, we consider an entry to belong
to a particular field if any of its photometric measurements has been
given that field number.

\subsection{Reflection halos}

The brightest stars in the \USNOB\ are surrounded not just by
diffraction spikes but by a thin circular ring or ``halo''.  This halo
is caused by internal reflections in the camera. Because this has a
geometric-optics rather than a physical-optics origin, the halo radius
is not a function of the wavelength of the imaging bandpass. This
means that spurious ``halo'' catalog entries can easily be present and
cospatial in multiple bands and thereby pass the \USNOB\ vetting
process (Figures~\ref{fig:skyPatchSource} and \ref{fig:skyPatch}).

Again, the spurious catalog entries can be identified by the patterns
they make on the sky.  Regularities include the following:
\textsl{(1)}~Halos are centered on extremely bright ($<7\,\mag$)
generating stars.  \textsl{(2)}~Halos have a circular or near-circular
shape.  \textsl{(3)}~Because they are very thin in the radial
direction, spurious halo entries have high two-dimensional density on
the sky.  \textsl{(4)}~The spurious halo entries are usually close to
making up full circles, and only rarely appear in just a fragment of a
circle.  These regularities permit a sensitive search.

\subsection{Other spurious entries}

In addition to the spikes and halos we address above, there are other
categories of spurious catalog entries with other origins, including
but not limited to the following: \textsl{(1)}~There are some lines of
entries from fortuitously coincident features
(scratches, trails, handwriting, and
other artifacts) on overlapping plates.  \textsl{(2)}~There are some
duplicate entries for individual stars in sky regions where two fields
overlap. These are cases in which individual stars detected in multiple
fields have not been correctly identified as identical.
\textsl{(3)}~There are quasi-spurious clusters of entries in and
around extended objects such as galaxies, nebulae, and globular
clusters.

We are doing nothing about any of these spurious features, in part because they
do not have regularities that lend themselves to computer-vision
techniques we employ in finding the previously mentioned defects.
They also represent a much smaller fraction of the \USNOB\
entries than the spurious entries from diffraction spikes and
reflection halos.

Of course the \USNOB\ contains also many entries that are in fact
compact galaxies rather than stars.  However, these entries are
\emph{not} spurious from our perspective, since compact
galaxies are as good as---or better than---stars for our \an\
astrometric calibration efforts, and most other astrometric calibration
tasks.

\section{Methods}

The Catalog we begin with is not the unmodified \USNOB, but rather the
\USNOB\ with the \Tycho\ Catalog \citep{hog00a} stars
re-inserted by us from the official \Tycho\ Catalog release.  We were
forced to perform this operation because in the official \USNOB\
release, the \Tycho\ Catalog stars were added in an undocumented
binary format.

\subsection{Diffraction Spikes}

We begin by dividing the Catalog into a fine healpix \citep{gorski05a}
grid, and projecting the entries in each healpixel onto planes tangent
to each healpixel's center. For each entry we calculate the average
$m$ of all magnitudes of all bands in which the entry has been
detected, and we find the union of all fields in which the entry is
present.

For each field present, we construct a ``profile'' of the field's
largest diffraction spikes, by overlaying the local neighborhoods of
the ten brightest stars in the field.  Given the regularities
discussed above, we can expect all spikes in each field to have the
same orientation. Therefore, each composite profile has one dominant
orientation, which is more apparent than in any single star's
neighborhood. To find each field's orientation, we first convert the
composite profile into polar coordinates, collapse the angles of each
point into a ${\pi\over 2}\,\rad$ range, and calculate a rough
histogram of the resulting angles.  The angle with the most densely
populated bin is used as an initial guess of the field's orientation,
which is then re-estimated using an iteratively reweighted least
squares (IRLS) fitting algorithm for robust M-estimation
\citep{hampel86}. The M-estimation is guaranteed to converge to an
estimate of the orientation that locally minimizes a total cost
$\sum_k \rho(e_k)$ where $e_k$ is the angular distance of entry $k$
from the estimated orientation.  We employ a Geman-McLure (GM) cost
function $ \rho(e_k) = e_k^2/(\sigma^2 + e_k^2) $, where $\sigma$ is
the initial guess of the root-variance of the angular width of a
spike.  This GM cost function replaces the standard least-squares cost
function $\rho(e_k) = e_k^2/\sigma^2$ and thereby downweights
outliers.  The resulting angle is a very robust and precise estimate
of the average orientation of all diffraction spikes present in the
field.

We iterate over fields, using our estimation of each field's dominant
orientation to rotate the entries present in each field such that the
diffraction spikes present become axis-aligned on average, making
their detection much easier. Because there is sometimes some
discrepancy between the position of the diffraction spike's generating
star and the center of the diffraction spike, we perform a robust
estimation of the centerpoint of the spike, just as we did in
estimating the orientations of the field profiles. With the
diffraction spike axis-aligned and zero-centered, we collapse all of
the entries in the neighborhood of the diffraction spike into a single
composite of all four ``corners'' (as if we were to convert the
neighborhood to polar coordinates, and collapse their angles into a
${\pi\over 2}\,\rad$ range), thereby reducing the four-part
diffraction spike to a single dense cluster of points.

We found a power-law approximation to the relationship between the
magnitude of the generating star and the angular extent of the
diffraction spike it generates among spurious entries.  This was found
by initially hand-labeling a small subset of the data, making a crude
fit to the hand-labeled data, then later refining the estimate using
the results of our algorithm.  Given the magnitude of a generating
star, we are able to use this relationship to estimate the angular
extent of the spike we would expect that generating star to produce.
As previously mentioned, the width of each spike is roughly
independent of the magnitude of the generating star, and is therefore
initialized to a constant value.  This estimate of the center and
extent of the spike is used to initialized a two-dimensional Gaussian,
which is then fit to the entries belonging to the diffraction spike
using iterated variance clipping at 2.5\,sigma.  What we construct is
not a traditional multivariate normal distribution, which would assume
that the data lies in an elliptical distribution, but is instead a
``rectangular'' distribution. That is, we consider an entry to be
within the Gaussian distribution if it is simultaneously within
2.5\,sigma of the width and 2.5\,sigma of the length of the
distribution.  When the Gaussian converges to its final parameters, we
take the rectangular area within 2.5\,sigma, and extend its range
towards the generating star to cover all entries between the area and
the generating star at the center of the spike.  If this area's
angular width, length, and position all pass a set of thresholds,
detailed later, we flag all of the entries within it (excluding the
generating star and any \Tycho\ stars, which we assume are not
spurious) as potential spurious entries.  If these entries pass a set
of thresholds (detailed below) they are marked as spurious.

\begin{figure}
\resizebox{\threewidth}{!}{\includegraphics{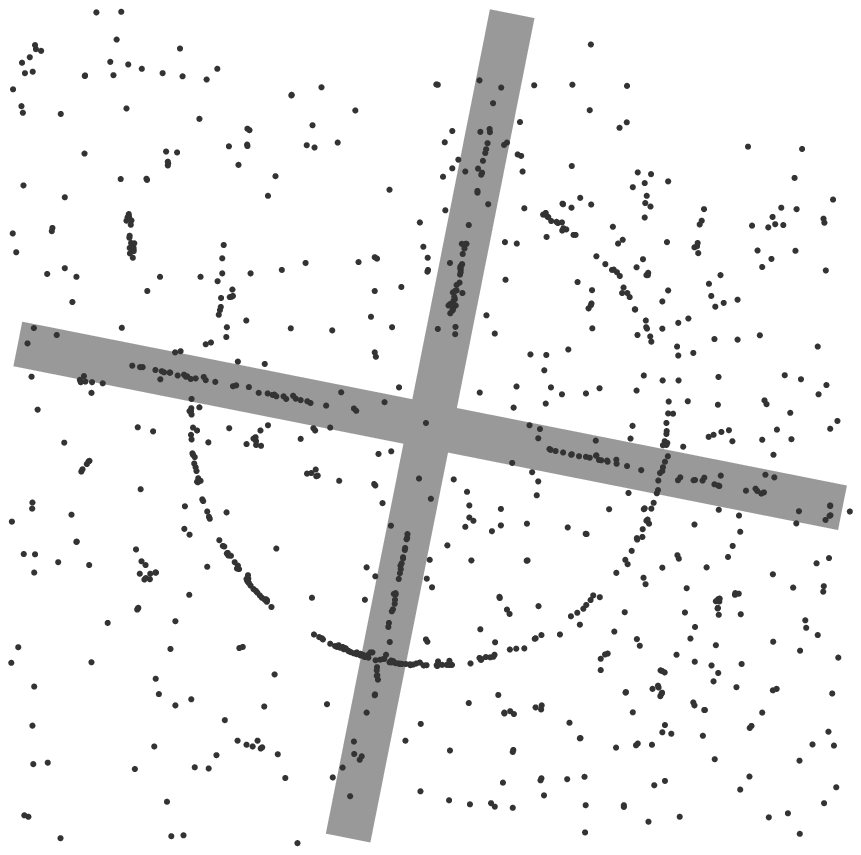}}%
\resizebox{\threewidth}{!}{\includegraphics{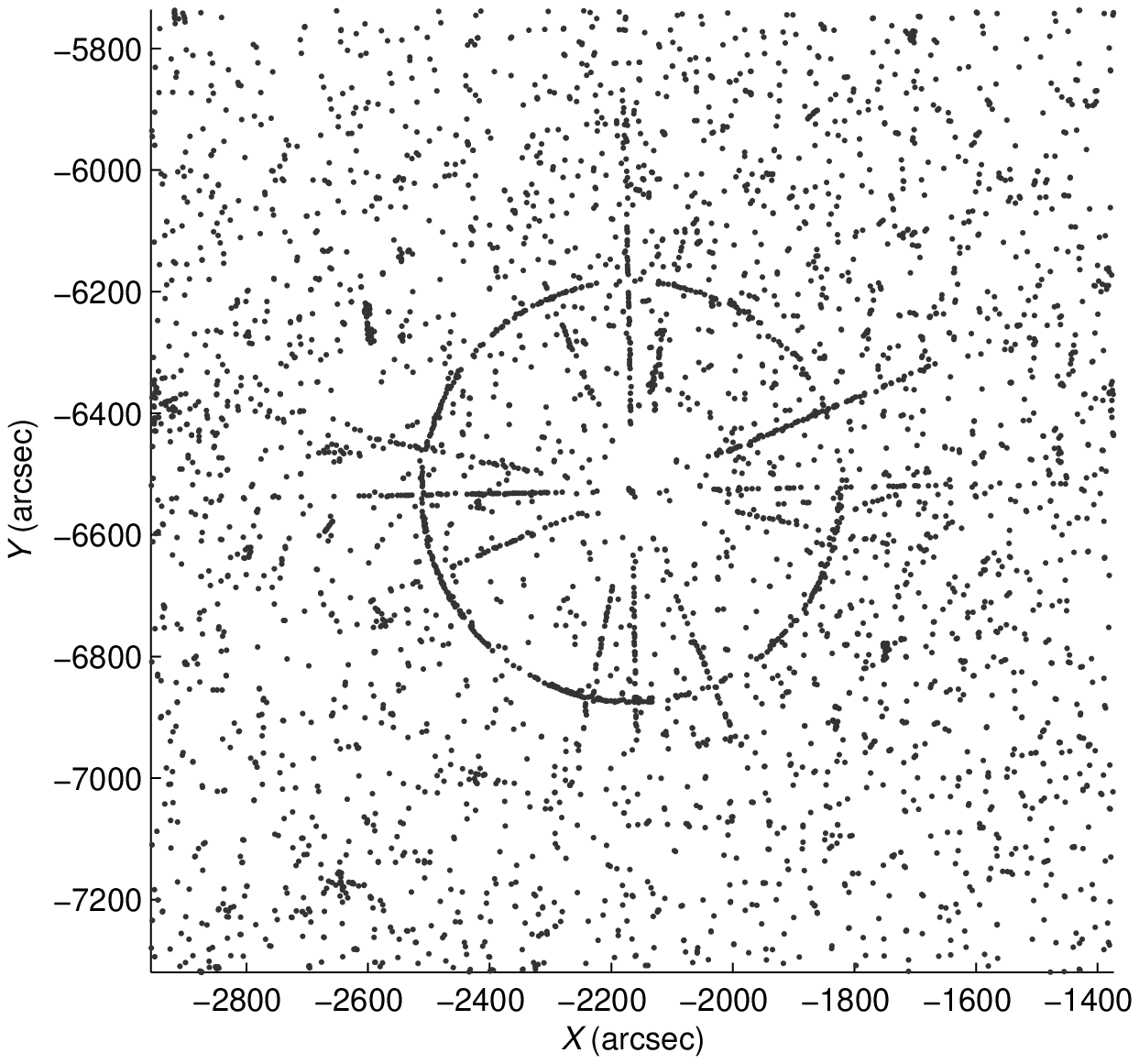}}%
\resizebox{\threewidth}{!}{\includegraphics{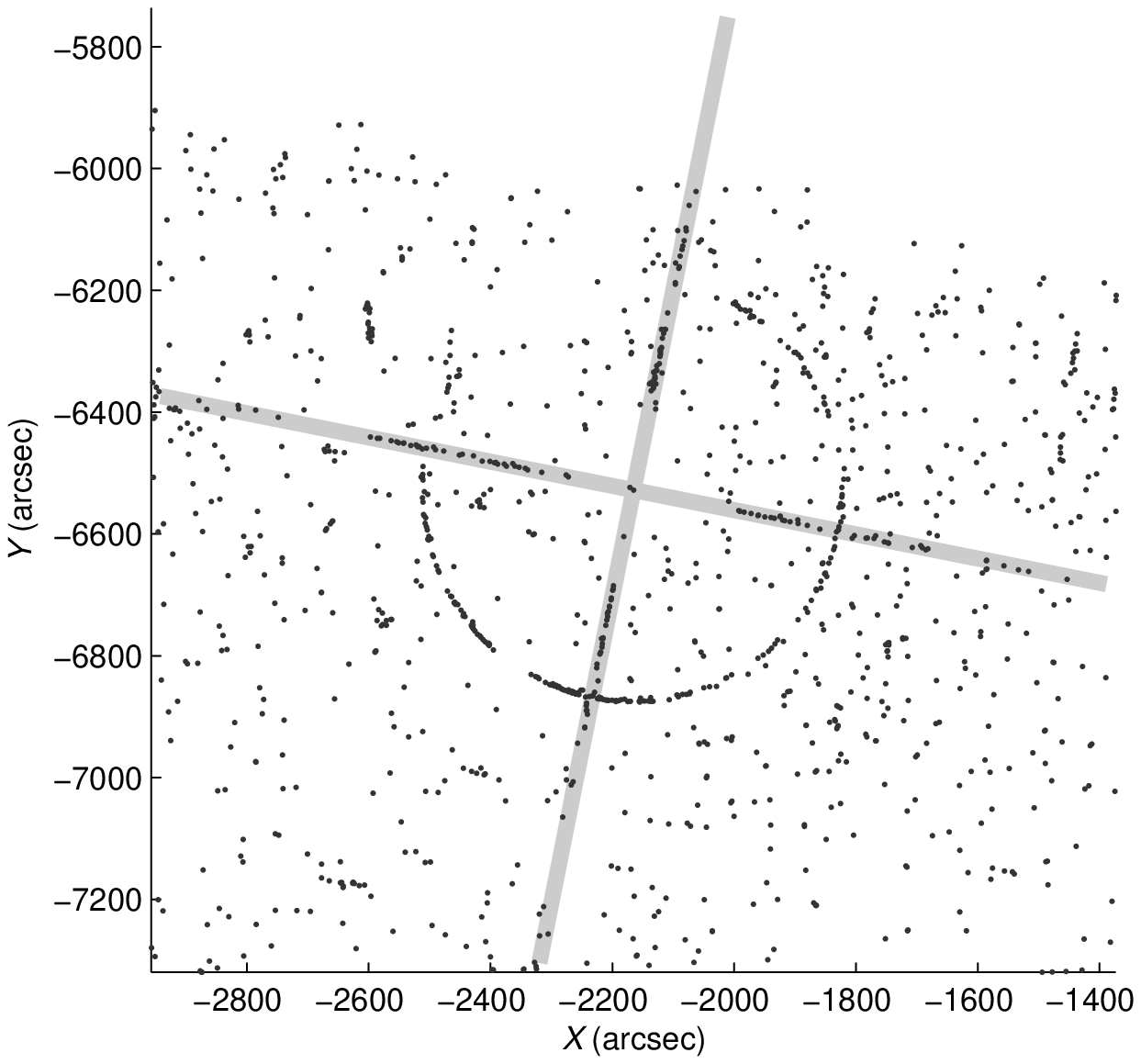}}
\resizebox{\threewidth}{!}{\includegraphics{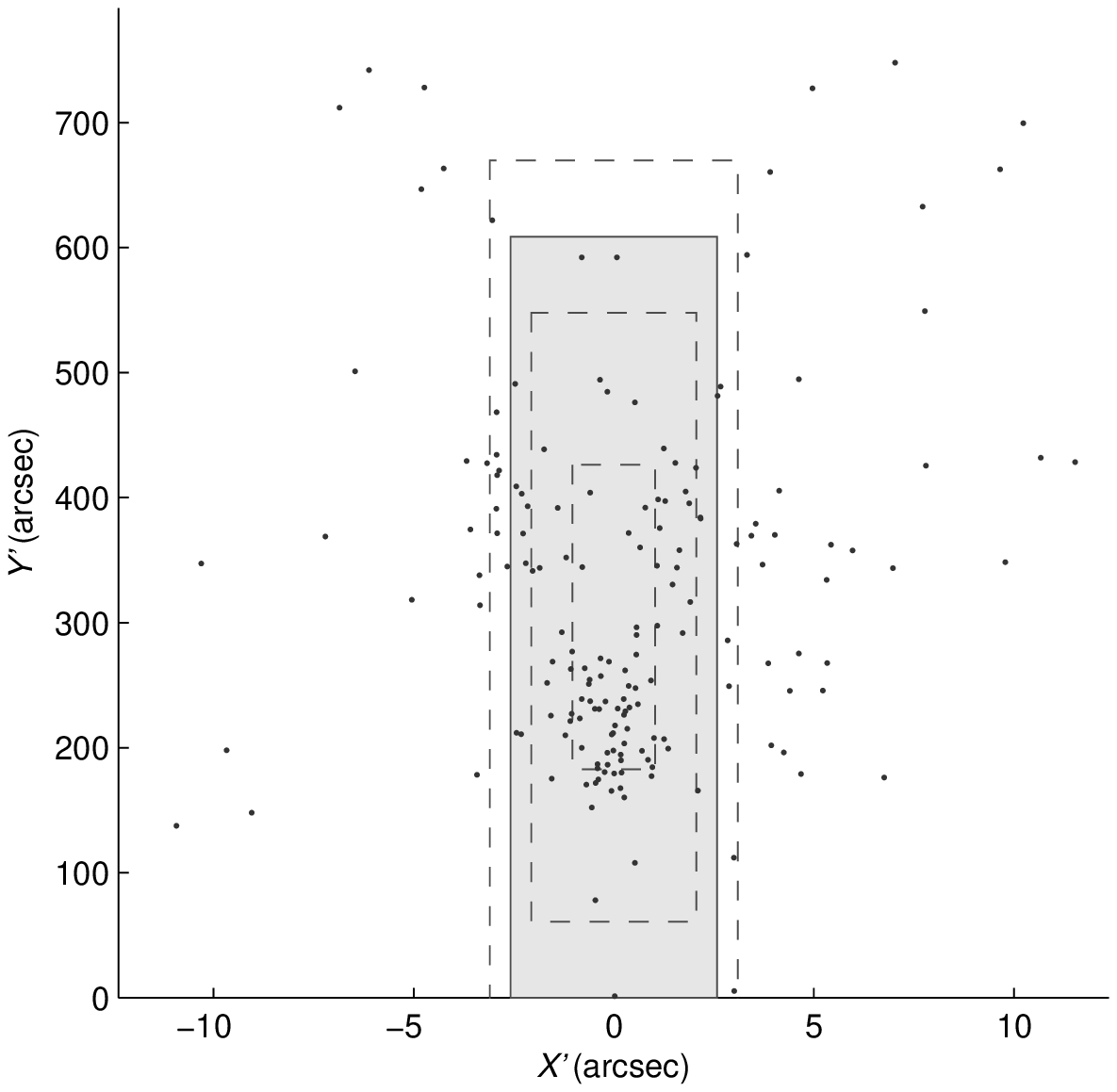}}%
\resizebox{\threewidth}{!}{\includegraphics{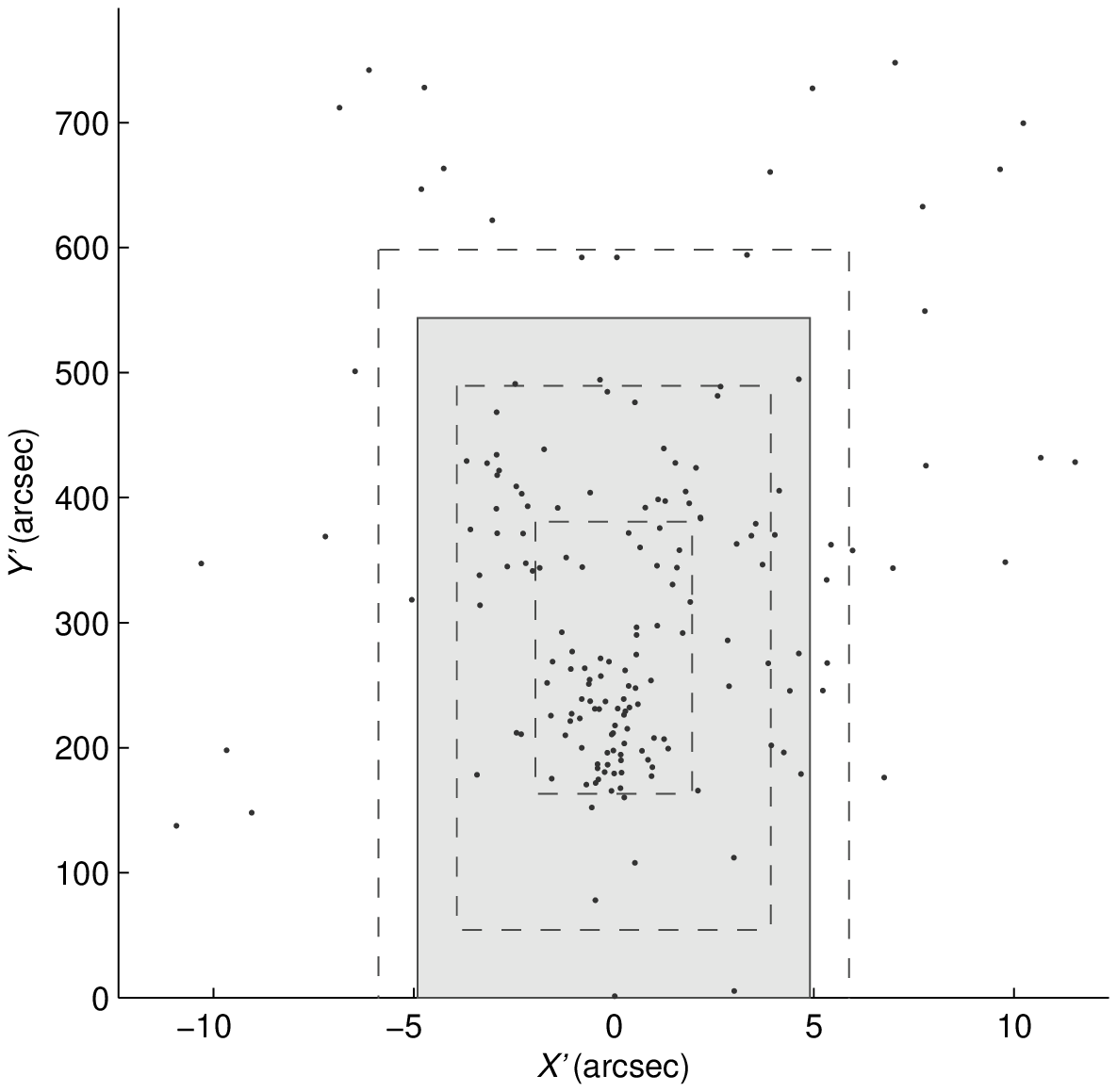}}%
\resizebox{\threewidth}{!}{\includegraphics{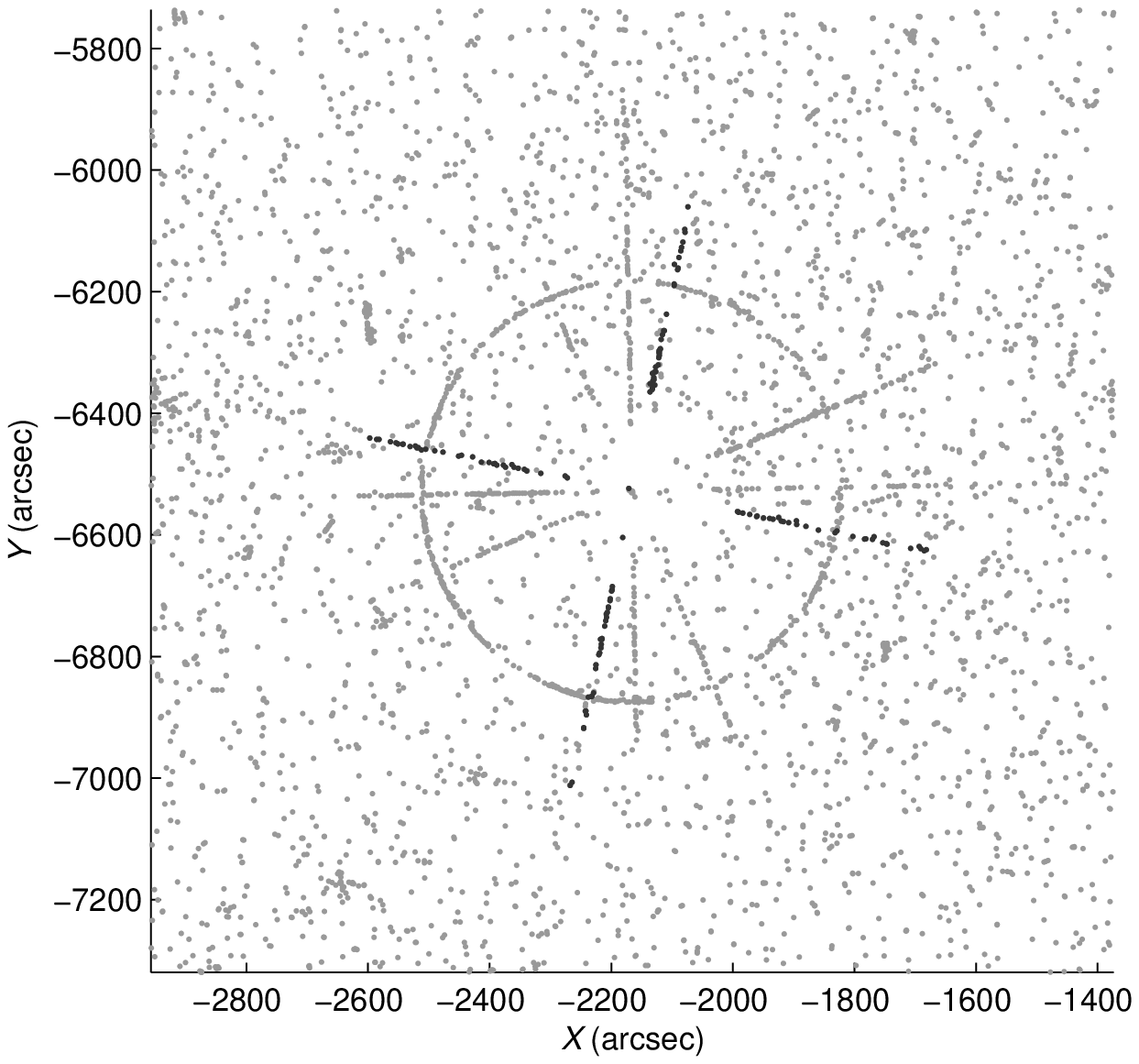}}
\caption{A narrative demo of finding a single spike within the patch of sky shown in Figure~\ref{fig:skyPatch}.
\textsl{Upper left panel:} The composite profile of the current
field, with the field's estimated orientation highlighted.
The profile is dominated by the spike
of the current generating star, which is the largest in the field.
\textsl{Upper center panel:} The neighborhood surrounding the 
current generating star, with all entries in all fields shown.
\textsl{Upper right panel:} The same neighborhood, with only entries
in the current field shown. The dominant orientation from this field's 
composite profile is highlighted.
\textsl{Lower left panel:} All four directions of the spike, collapsed
into one ${\pi\over 2}\,\rad$ profile.
The dashed lines outline the areas encompassed at 1, 2, and 3 times the
root-variance of the Gaussian used to initialize the variance clipping.
The solid rectangle is the area encompassed at 
2.5\,sigma, which is threshold for flagging entries. The solid rectangle
is extended all of the way to the bottom, as it is assumed that all of the 
spike profile between the spike cluster and the generating star is also
spurious.
\textsl{Lower center panel:} The same profile, after the Gaussian has
been fit using variance clipping. The solid rectangle shown here
is the final area we use for determining if entries are flagged.
\textsl{Lower right panel:} The neighborhood surrounding the generating
star with all entries shown, and with the newly-flagged spike entries darkened.
The diffraction spikes in the other orientations come from different fields and
are flagged by later passes of the algorithm.
\label{fig:demoRun}}
\end{figure}

The algorithm is depicted in Figure~\ref{fig:demoRun}.

\subsection{Reflection halos}

Once all spurious catalog entries attributed to spikes are found and
temporarily removed (such entries disturb the results of the halo
detection algorithm), we search the remaining catalog for halos.  This
process is similar to the process of searching for diffraction spikes:
We divide the Catalog into a fine healpix grid and process each grid
cell independently.  We project the entries in each grid cell onto a
plane tangent to the cell's center.  Next, we examine each star
brighter than $7\,\mag$, and attempt to find and eliminate halos that
it has generated.  Since the radius of each halo is not dependent on
the magnitude of its generating star, the size of the neighborhood we
search is constant.

We convert each neighborhood into polar coordinates centered at the
generating star, and calculate a histogram of the radii of all entries
in the neighborhood. This simple count of the number of stars present
at different radii is used to generate a more informative histogram of
the densities of stars at each radius. Our initial guess of the radius
of the halo is whichever coarse bin is the most dense.

With this estimate of the radius of our halo, and with a constant as
our initial estimate of the radial width of the halo, we construct a
one-dimensional Gaussian and again robustly fit the position and width
of the Gaussian using variance clipping at 2.5\,sigma.  Once the
re-estimation has converged, we check that our resulting values for
the variance of the width are reasonable ($<3\,\arcsec$), and if so,
we label all entries within 2.5\,sigma of the Gaussian as potentially
spurious. Again, if these entries pass another set of thresholds, they
are marked as spurious.

Because one generating star may produce multiple halos, we search each
generating star, and remove each salient halo we find, until we fail to 
detect any new halo that passes our thresholds.

\subsection{Parameters of the Algorithms}

By necessity the algorithms have a number of free parameters.  Some of
these are measurements of diffraction-spike and reflection-halo
configurations, derived from quantitative analyses of the properties
of the spurious entries, while others are additional conservative
constraints, applied to ensure that the spurious entries appear to be
correctly identified on visual inspection of the results.

In addition to the parameters that specifically apply to the spike and
halo identification algorithms, we somewhat arbitrarily chose to work
in a $\Nside=9$ healpix grid; there are $12\times 9\times 9=972$
healpixels.  We set all variance-clipping thresholds to 2.5\,sigma,
and when we define regions by variance clipping we make them
2.5\,sigma in half-width.

\subsubsection{Measured Spike Parameters}

\begin{itemize}
\item We search for diffraction spikes generated by stars brighter
than $13\,\mag$.  Bright stars tend to produce large diffraction
spikes containing many spurious entries, while dimmer stars produce
small diffraction spikes containing few, and potentially ambiguous,
spurious entries.  When we extended our search to stars brighter than
$15\,\mag$, we found that the proportion of falsely labeled spurious
entries increased dramatically.  Our decision to restrict to
$<13\,\mag$ is further supported by the second panel of
Figure~\ref{fig:spuriousStats}, which shows that generating stars at
$>13\,\mag$ have mean number of entries per spike less than four,
which means that most will contain too few entries to be accepted.

\item Our initial estimate of the angular length $\ell$ of a
diffraction spike given the magnitude $m$ of its generating star is
$\ell = 3500 \times 1.53^{-m}\,\arcsec$; see
Figure~\ref{fig:spikeProperties}.  This estimate initializes a
refinement by iterated variance clipping and therefore does not
strongly affect our results.  In detail this relationship between
length and magnitude depends on band, exposure time, and data quality,
and is is therefore different for every plate; but since we use it
only as an initialization, those details do not substantially affect
our results.

\item Our initial estimate of the angular width of a spike is
$1\,\arcsec$.  This also initializes a refinement by iterated variance
clipping and also has little effect on our results.

\item We define the ``reasonable'' width of a diffraction spike to be
three times the initial estimate of $1\,\arcsec$.  If the adaptive
fitting process produces a width larger than this, the candidate spike
is rejected.
\end{itemize}

\subsubsection{Additional Spike Constraints}

\begin{itemize}
\item The size of the local neighborhood constructed around each spike
is $2.5$ times the initial estimate of the spike's size.  This limits
the catalog entries considered in the subsequent analysis, though the
effect on our results is minimal.

\item We required each spike to have entries in at least $2$
of the $4$ spike regions.

\item We required the total area within the four spike regions
to be at least as dense in Catalog entries as the surrounding area.
\end{itemize}

\subsubsection{Measured Halo Parameters}

\begin{itemize}
\item We search for halos around generating stars brighter than
$7\,\mag$.  Our experiments have shown that halos do not appear around
stars dimmer than this.

\item We discard any halo whose radius is outside the range of $240$ to
$410\,\arcsec$.  Direct inspection of the catalog shows that
reflection halos rarely appear outside of this range.

\item Our initial estimate of the standard deviation of the radial
width of a halo is $1.8\,\arcsec$.  This is approximately the average
value to which our variance-clipping fitting algorithm converges.

\item We discard any halo for which our variance-clipping fitting
algorithm computes a radial width larger than $4.5$ times the initial
estimate.
\end{itemize}

\subsubsection{Additional Halo Constraints}

\begin{itemize}
\item Each halo must contain at least $25$ catalog entries.

\item The density of catalog entries in each halo annulus
must be at least $1.25$ times the density of the area near the halo.

\item There must be entries present in the halo annulus every
${\pi\over 4}\,\rad$. This forces all detected halos to be fully
circular, rather than just fragments of circles. More importantly,
this requirement prevents the false detection of halos near the edges
of healpixels, which would otherwise happen very often. Unfortunately,
this requirement prevents us from detecting any halo near the edges of
a healpixel.
\end{itemize}

\subsection{Limitations}

Limitations of our procedures include the following.
\begin{itemize}
\item The algorithm assigns hard labels to indicate that an entry is
spurious.  A future version of the algorithm could assign an
assessment of our \emph{confidence} that an entry is spurious.

\item The algorithm processes each healpixel independently, and we
have not included a buffer region around the edges of the healpixels,
so there are minor edge effects: the algorithm is less likely to
detect spurious entries near the healpixel boundaries.  We expect this
to affect roughly $0.4\,\percent$ of the diffraction spikes and
$3.5\,\percent$ of the reflection halos.

\item The algorithms are highly specialized to the typical data in the
\USNOB. If a small fraction of the data in the Catalog come from some
telescope with, for example, three rather than four supports for the
secondary, or very different internal reflections, the algorithms we
use would not detect the spurious features in those data.

\item There are many hard settings of parameters, as discussed above.
Most of these are either just initializations for iterative procedures
or else set manually after an analysis of the data, but more
experimentation could have been performed if we had a substantial data
set in which the spurious entries had been reliably identified in
advance.

\item Sometimes a diffraction spike that exists in multiple fields is
detected in a field whose orientation does not match the spike's orientation
as well as some other field. The is because the order in which we search
each field is arbitrary; we flag a detected diffraction spike upon
it's first successful detection. This usually results in a detected
diffraction spike with an unusually wide angular width. Though this happens
frequently, its overall effect on the fidelity of our results is small.
A better solution would be to remove spikes in non-increasing order of
their resemblance to our model of a diffraction spike.

\item We ought never consider as a generating star any star that was
marked spurious in the analysis of a brighter generating star. We
don't currently enforce this, and it may produce some incorrect
identification of spurious entries.

\item Many of these limitations could be overcome if we constructed
a complete generative model of
diffraction spikes and halos.  This would allow us to ``score'' potential spurious
detections with something approaching a \emph{probability} that they are spurious,
rather than simply cut at hard thresholds. This could also improve the fidelity of our results, 
by allowing us to increase our statistical requirements of some parameters
of our generative model when a detected spike or halo fails to fit other
parameters. For example, if a possible halo appears at an uncommon radius,
a proper generative model would effectively put a stronger constraint on
other properties (such as the density of entries in the halo annulus)
in order for the entries to be marked as spurious with high probability.
Done well, this approach could also allow us to reduce the number of 
individual requirements we require of each detected spike and halo.
This would be aided by a set of hand-labeled spikes and non-spikes, with
which we could tune the generative model --- or which we could use as
input to some kind of discriminator which would tune the model automatically.

\end{itemize}

\section{Results}

The number of entries flagged as spurious on diffraction-spike grounds
is \numSpikes\ (\percentSpikes\ of the \USNOB) and on halo
grounds is \numHalos\ (\percentHalos).  Our grounds for declaring an
entry spurious are conservative in the sense that a spike or halo is
only treated as being detected if it passes a set of statistical
thresholds.

The method works by marking as spurious all \USNOB\ entries in a set
of finite regions of the sky, with those sky regions adaptively fit to
the observed diffraction spike and reflection halo features.  Because
the total solid angle removed is non-zero, we expect some of the
entries we mark as spurious to in fact correspond to real sources. We
can estimate this in a representative healpixel: Healpixel 0 contains
$299573$ \USNOB\ entries; we flag as spurious $7924$ entries within a
set of regions comprising $1.5\times 10^{-5}\,\ster$ ($0.12\,\percent$
of the healpixel); we expect therefore some 300 of these to correspond
to real stars.  We tested this hypothesis with the
\TWOMASS\footnote{http://www.ipac.caltech.edu/2mass/}.  In this
healpixel there are $81089$ entries, of which we would expect
$\sim100$ to lie in the spurious area we've removed.  We find that
$82$ \TWOMASS\ entries match to a spurious \USNOB\ entry and no
non-spurious \USNOB\ entry, consistent with what we would expect
assuming a uniform distribution of \TWOMASS\ entries over the
healpixel. This count is probably an overestimate, because there are
some diffraction artifacts in the \TWOMASS\ that are similar to those
in \USNOB.  Our marking of spurious entries is aggressive in this
sense; as we noted in the Introduction, this is because for our
scientific purposes we require a catalog as clean of spurious entries
as possible.

\begin{figure}
\resizebox{\twowidth}{!}{\includegraphics{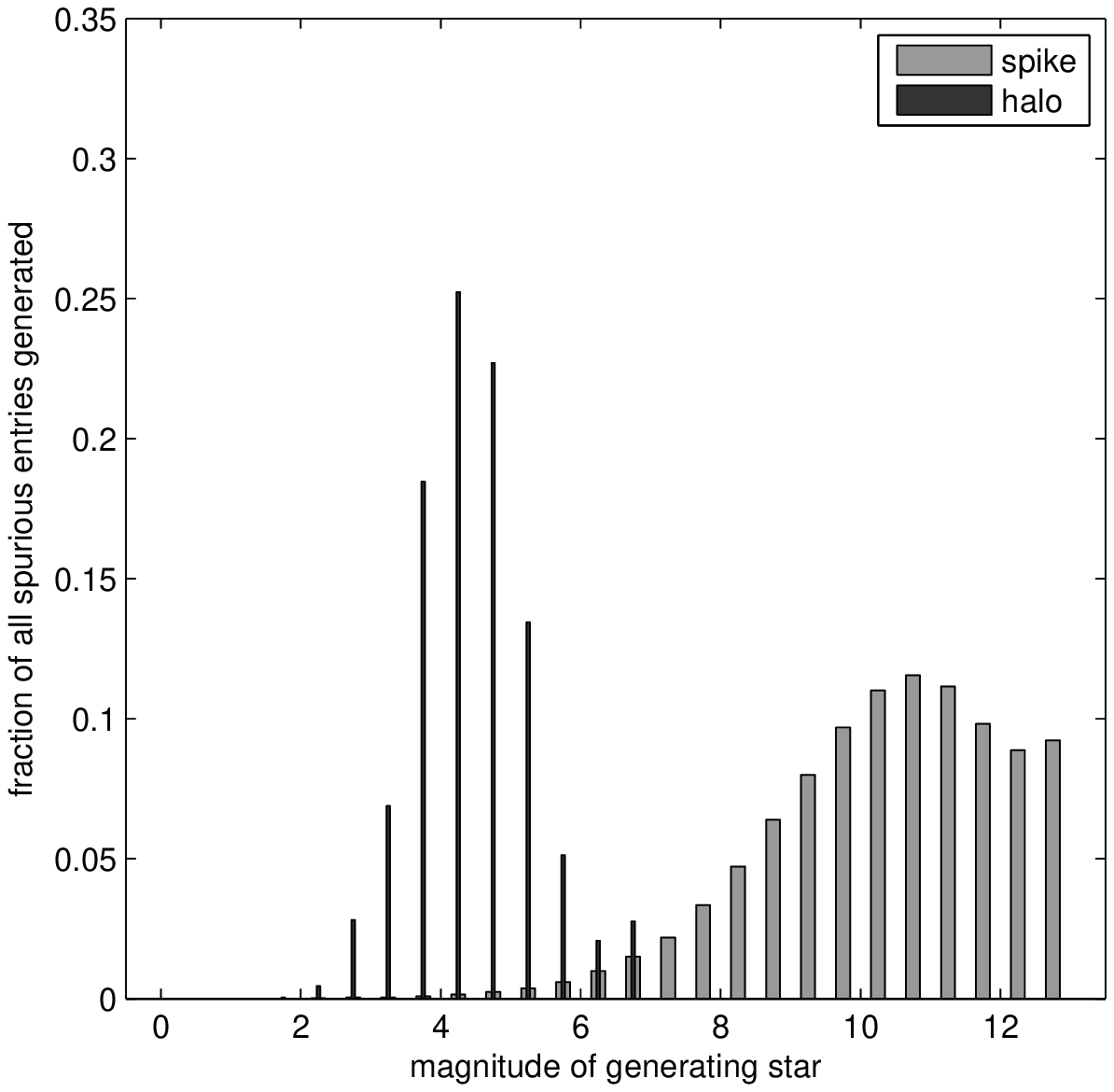}}%%
\resizebox{\twowidth}{!}{\includegraphics{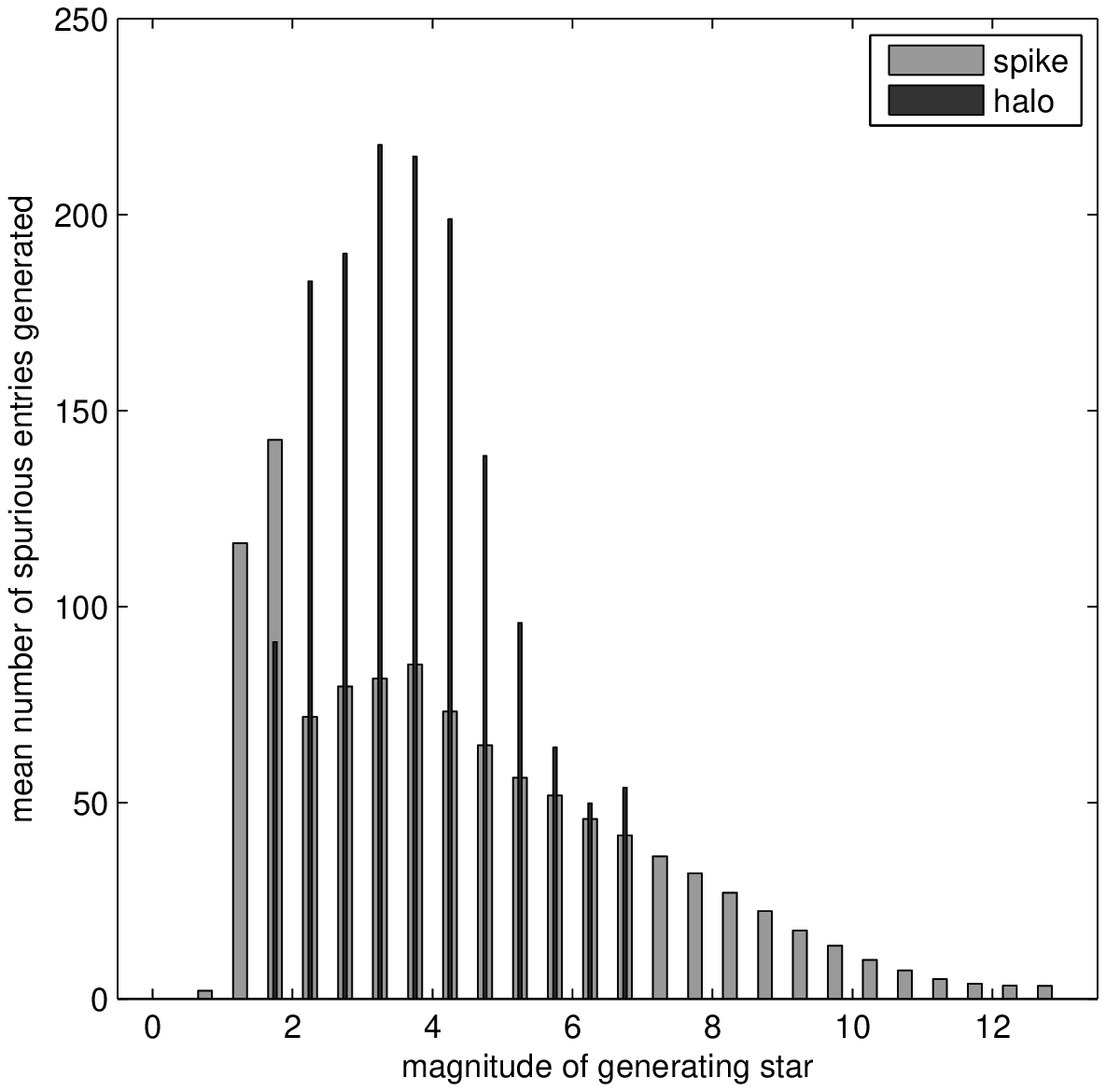}}\\
% dstn asks: can you make the legend show the dark thin bars as a dark thin patch?
% (right now it only shows the different in darkness, not of thickness)
% jon replies: Yeah, no, I can't, or not easily, at least. It's on my queue.
\caption{Statistics of spurious entries.
\textsl{Left panel:} The fraction of all entries marked as spurious
as a function of generating star magnitude.
\textsl{Right panel:} The mean number of entries marked as spurious
per generating star as a function of generating star magnitude.%
\label{fig:spuriousStats}}
\end{figure}

\begin{figure}
\resizebox{\twowidth}{!}{\includegraphics{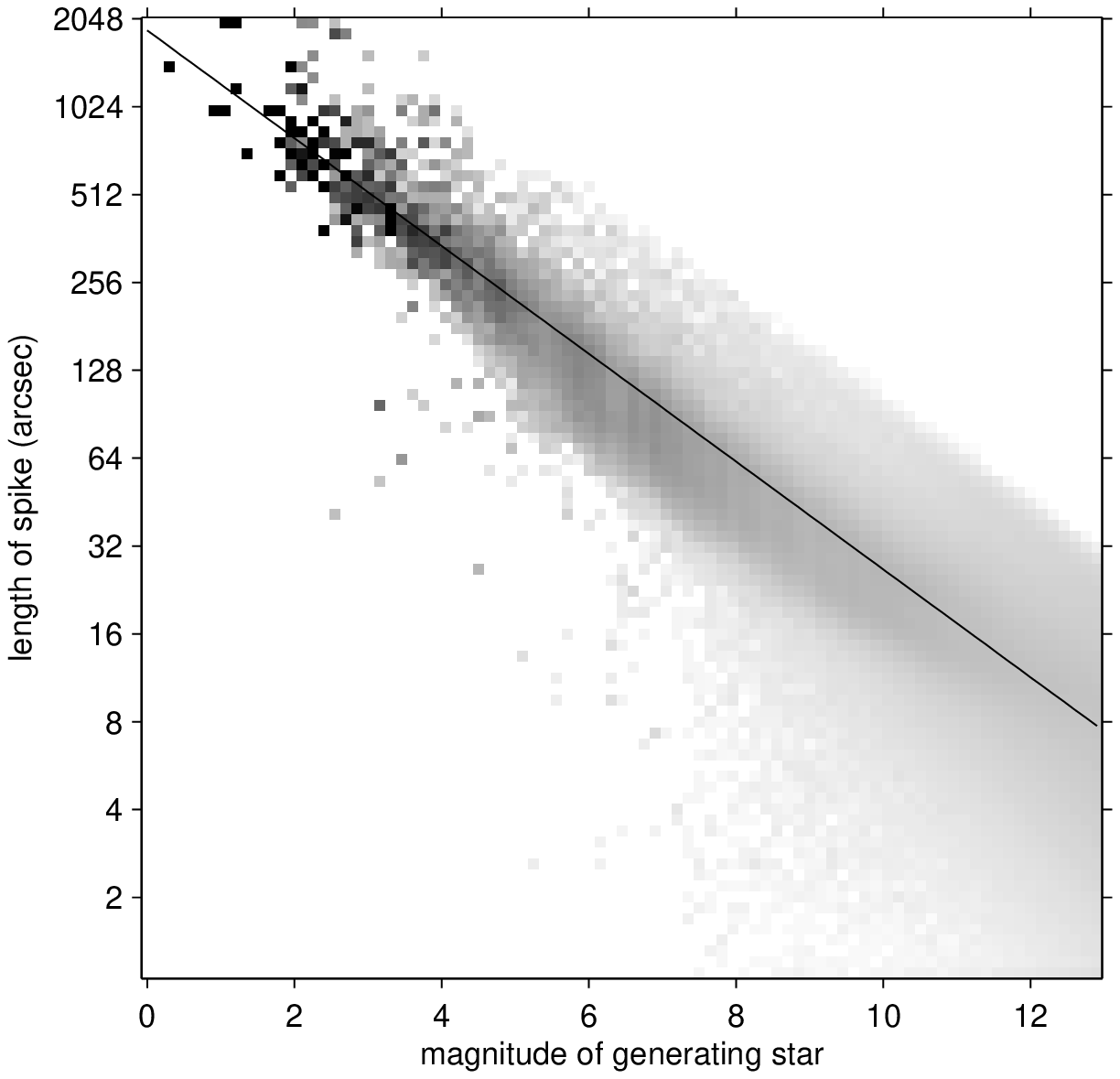}}%
\resizebox{\twowidth}{!}{\includegraphics{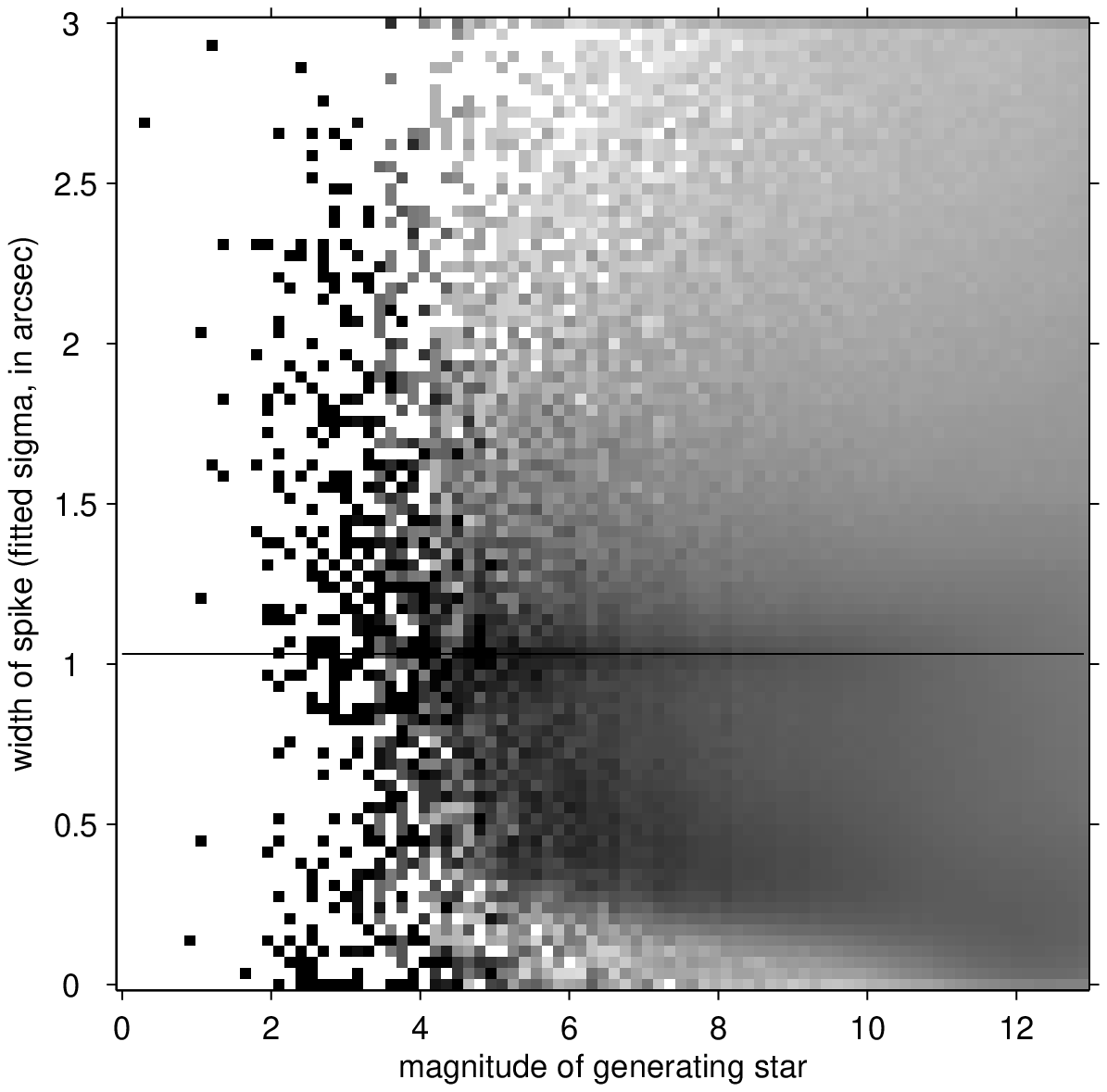}}\\
\resizebox{\twowidth}{!}{\includegraphics{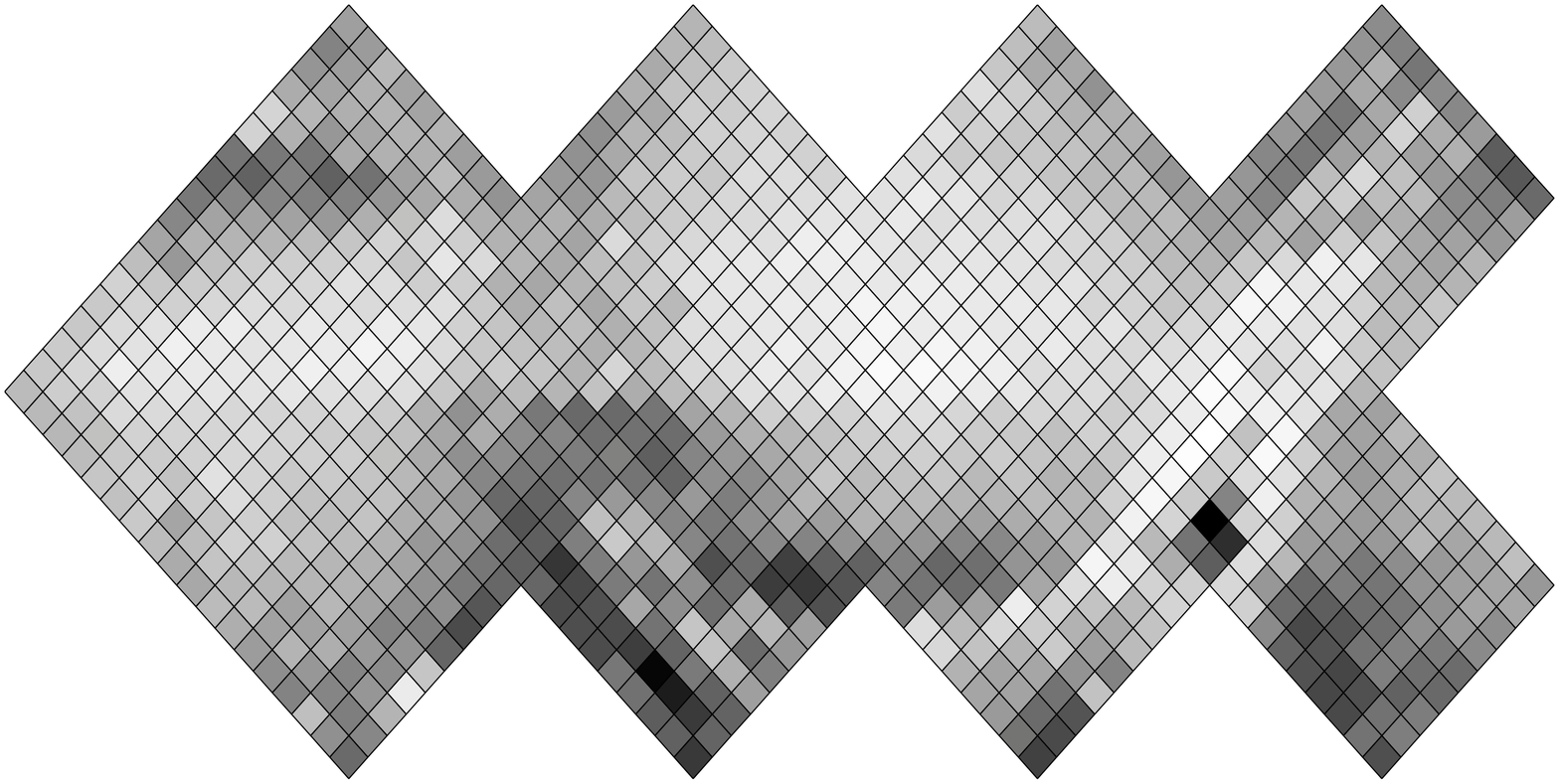}}%
\resizebox{\twowidth}{!}{\includegraphics{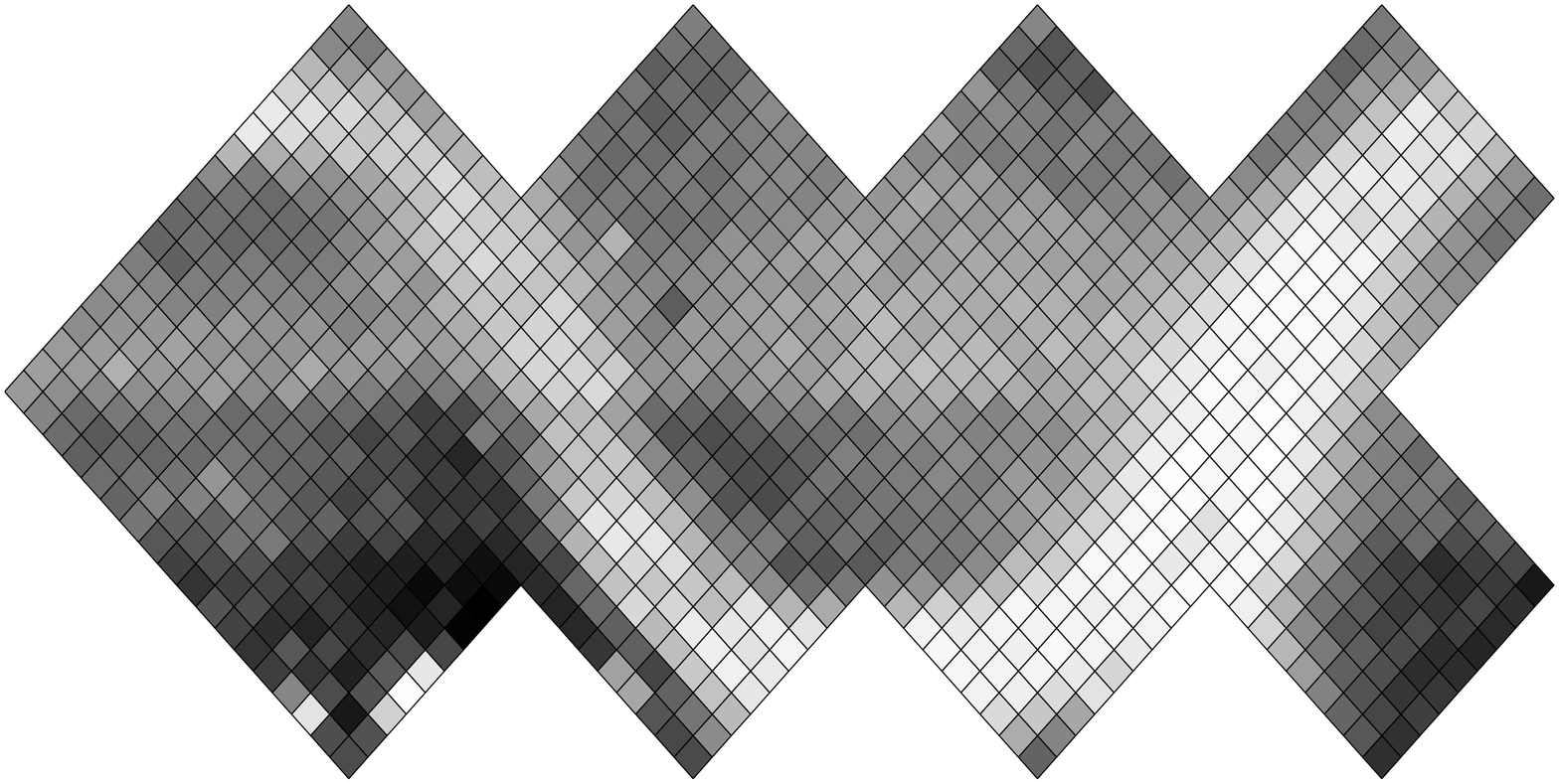}}
\caption{Regularities of spurious catalog entries in the \USNOB\
identified as caused by diffraction spikes.  \textsl{Top-left
panel:}~Two-dimensional histogram showing the adaptively fit radial
lengths of the spikes, found by iterative variance-clipping, as a function
of generating star magnitude.  Each vertical column in the histogram
is independently normalized.  The solid line shows the
value used to initialize the adaptive fitting.  \textsl{Top-right
panel:}~Similar two-dimensional histogram but showing the adaptively
fit widths of the spikes, as a function of generating star magnitude.
The solid line shows the initial value. \textsl{Bottom-left
panel:}~The two-dimensional solid-angular density on the sky of
spurious entries identified as parts of diffraction spikes as a
function of sky position (shown as an ``unwrapped'' $\Nside=9$ healpixel
grid). The darker a healpixel is, the more spurious ``spike'' entries it contains.
\textsl{Bottom-right panel:}~The same, but shown relative to
the number of catalog entries in that healpixel. In the two sky
density plots, a North--South asymmetry is visible, as well as the
Galactic plane.
\label{fig:spikeProperties}}
\end{figure}

\begin{figure}
\resizebox{\twowidth}{!}{\includegraphics{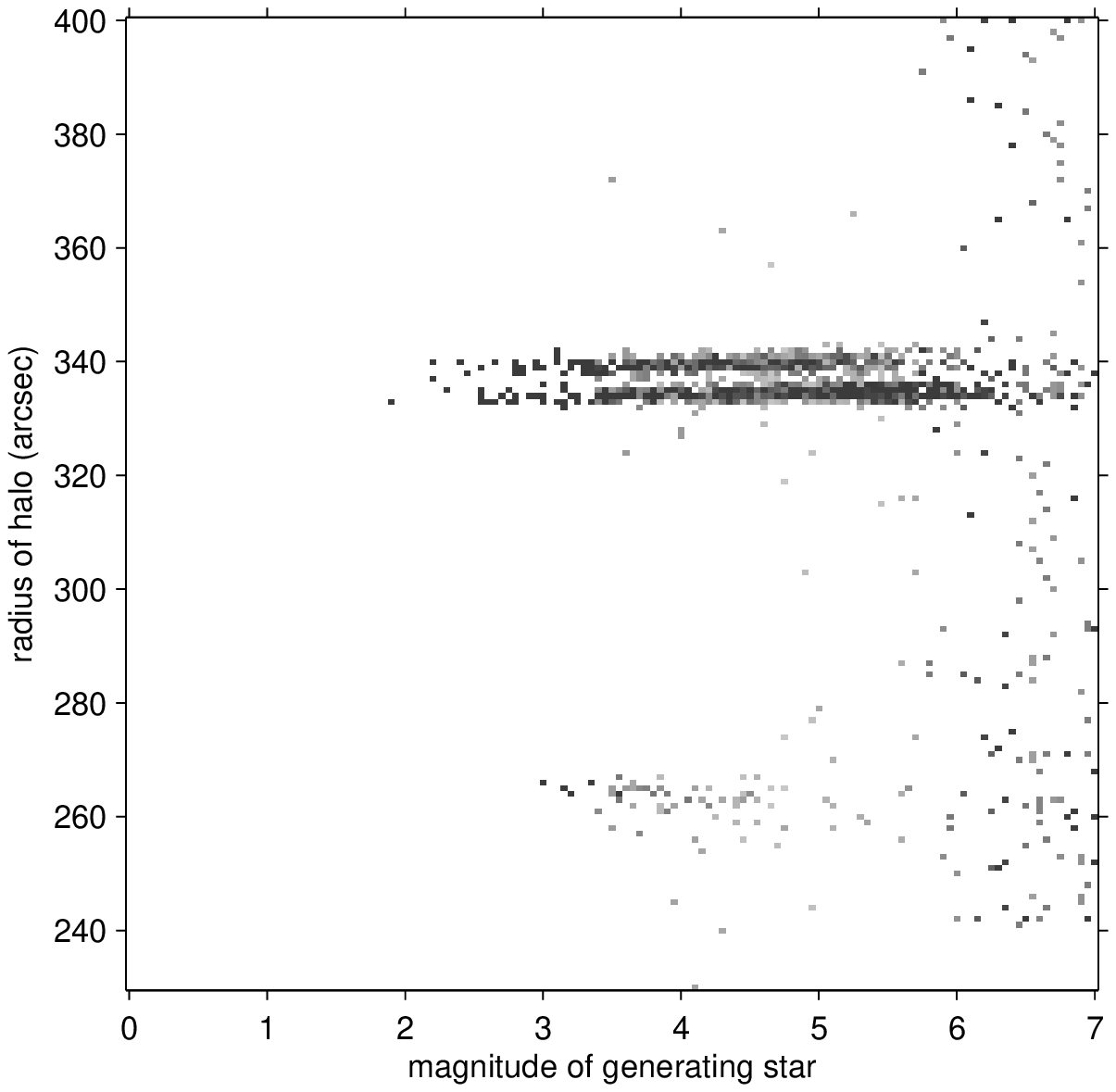}}%
\resizebox{\twowidth}{!}{\includegraphics{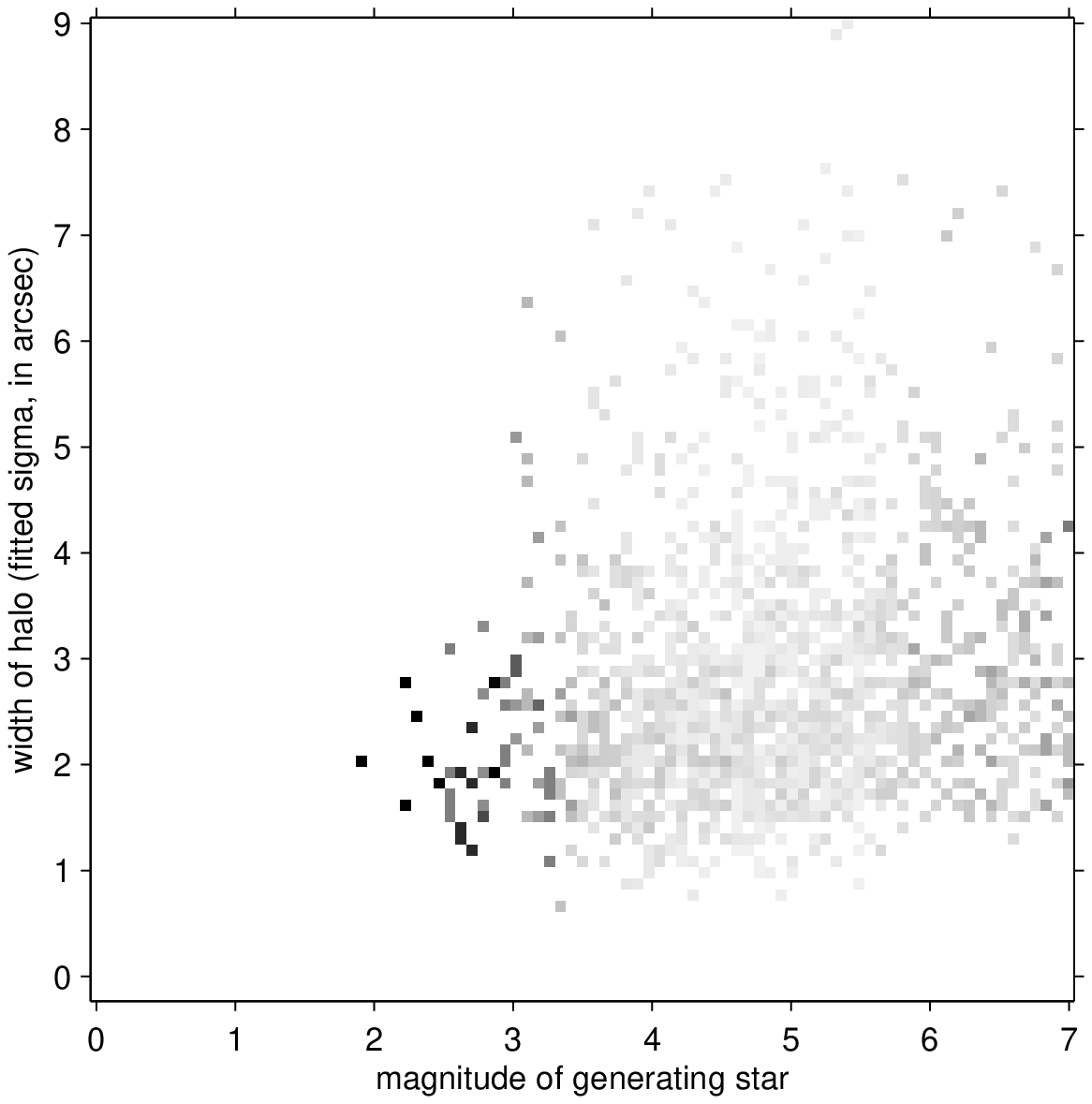}}\\
\resizebox{\twowidth}{!}{\includegraphics{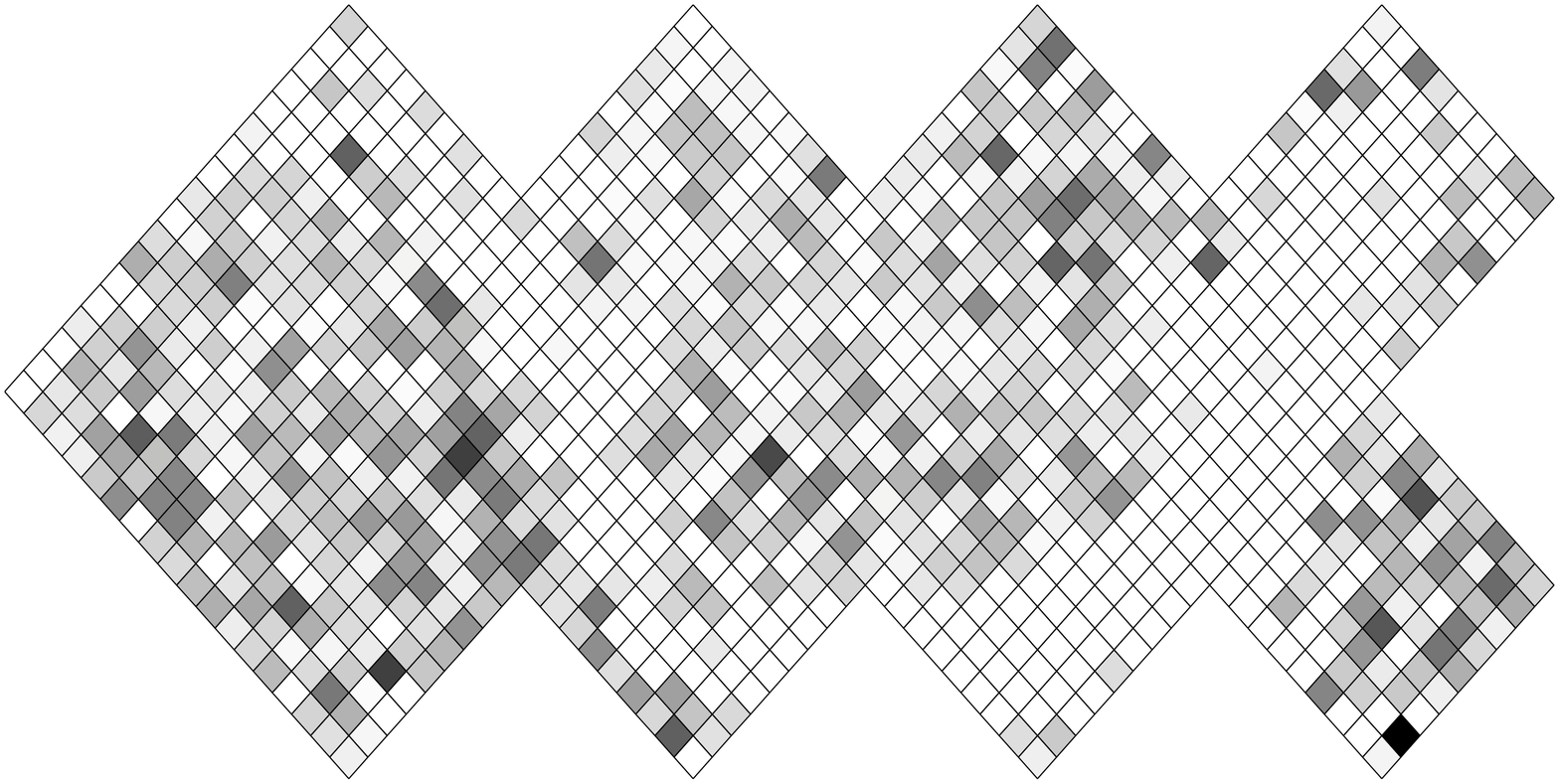}}%
\resizebox{\twowidth}{!}{\includegraphics{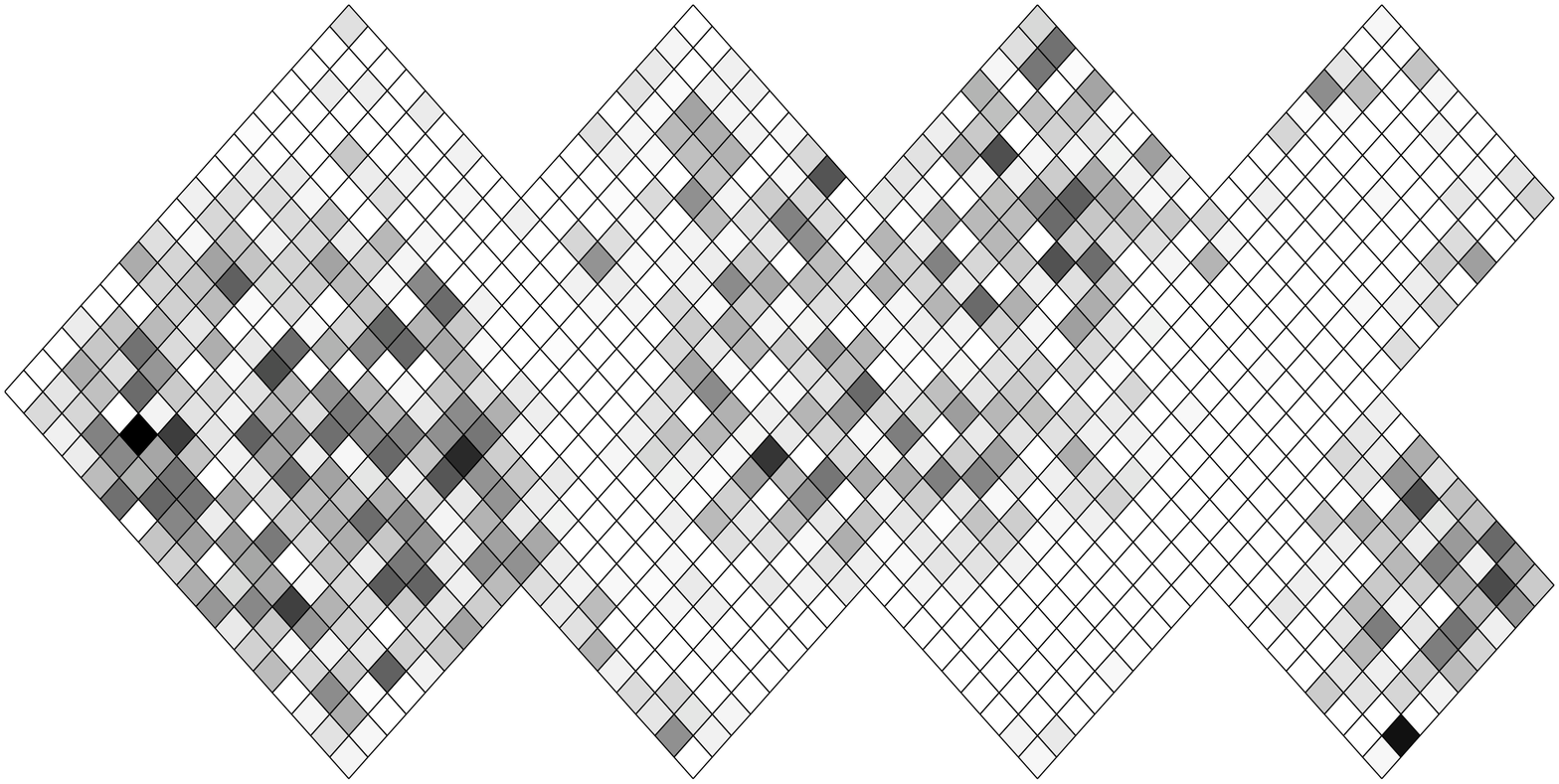}}
\centerline{\resizebox{\onewidth}{!}{\includegraphics{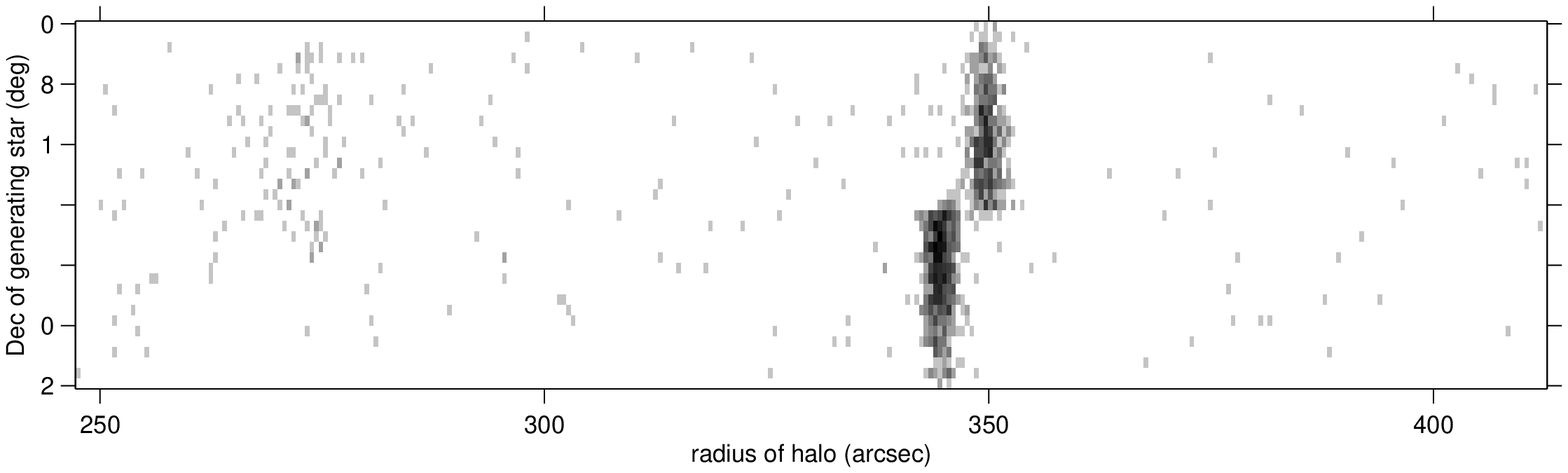}}}
\caption{Regularities of spurious catalog entries identified as caused
by reflection halos.  \textsl{Top-left panel:}~Two-dimensional
histogram showing adaptively fit reflection-halo radii as a function
of generating star magnitude.  Each vertical column of the histogram
has been independently normalized.  \textsl{Top-right panel:}~Similar
two-dimensional histogram but showing the adaptively fit widths of the
halo annuli.
\textsl{Middle-left panel:}~Two-dimensional solid-angular density on
the sky of spurious entries identified as parts of reflection halos.
The darker a healpixel is, the more spurious ``halo'' entries it contains.
\textsl{Middle-right panel:}~The same but shown relative to the number
of catalog entries in that healpixel.
\textsl{Wide bottom panel:}~Two-dimensional histogram showing that
each of the two principal halo radii is in one hemisphere of the sky.
\label{fig:haloProperties}}
\end{figure}

Properties of the spurious entries we have identified are shown in
Figures~\ref{fig:spuriousStats}, \ref{fig:spikeProperties}, and
\ref{fig:haloProperties}, including the numbers and fractions of
spurious entries as a function of generating star magnitude, and
distributions of spikes and halos in size and on the sky.  These
figures show a number of important regularities, for example that
brighter stars have larger diffraction spikes (as expected), that the
widths of the spikes is not a function of generating star magnitude
(also as expected), and that both the number of spurious entries and
our ability to robustly detect them are functions of sky position
(mainly because of the Galactic Plane).
Figure~\ref{fig:haloProperties} shows that there are two different
dominant halo radii, one for the North and one for the South;
presumably this indicates differences in the hardware used for each
hemisphere.

\begin{figure}
\resizebox{\twowidth}{!}{\includegraphics{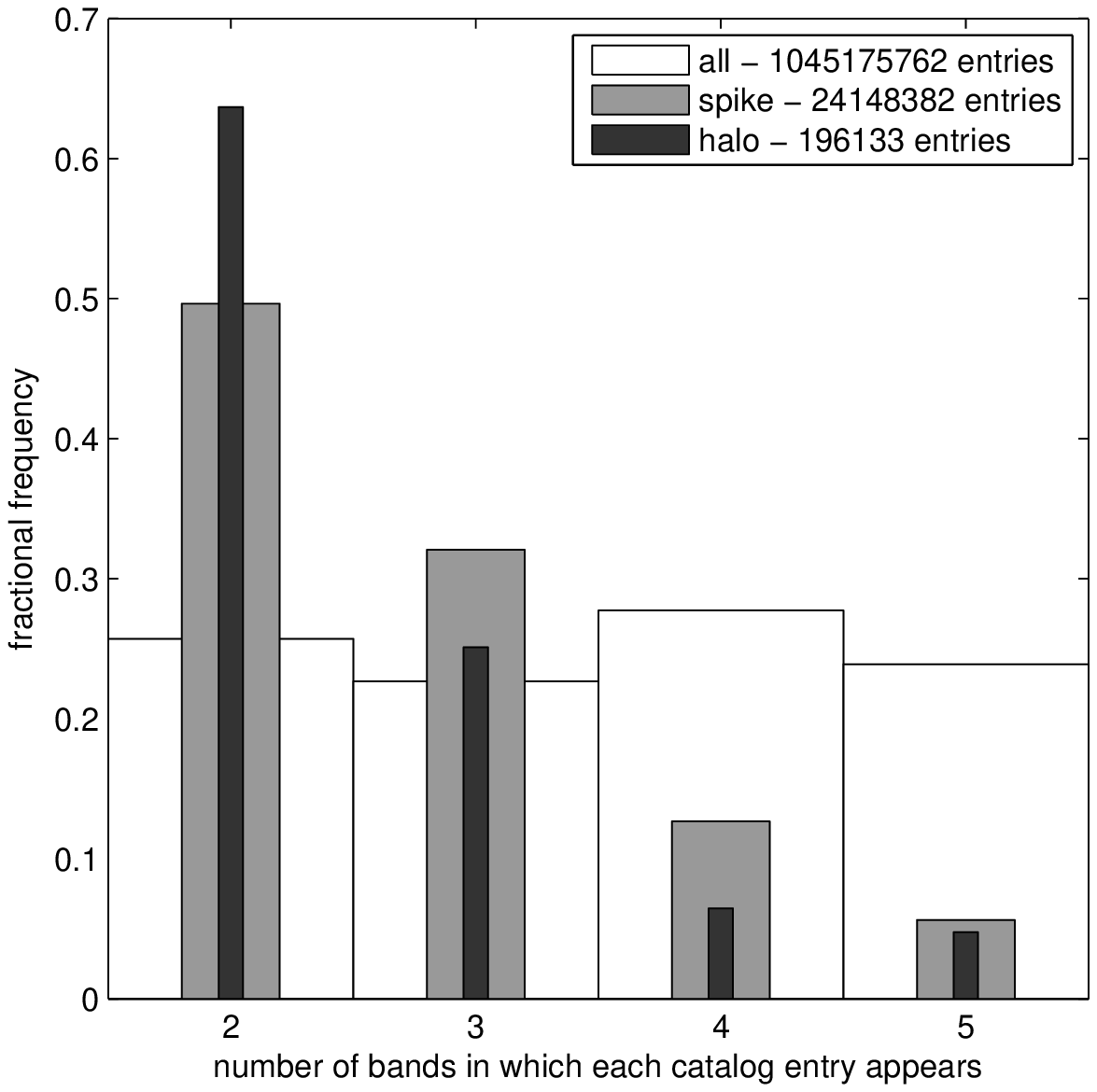}}%%
\resizebox{\twowidth}{!}{\includegraphics{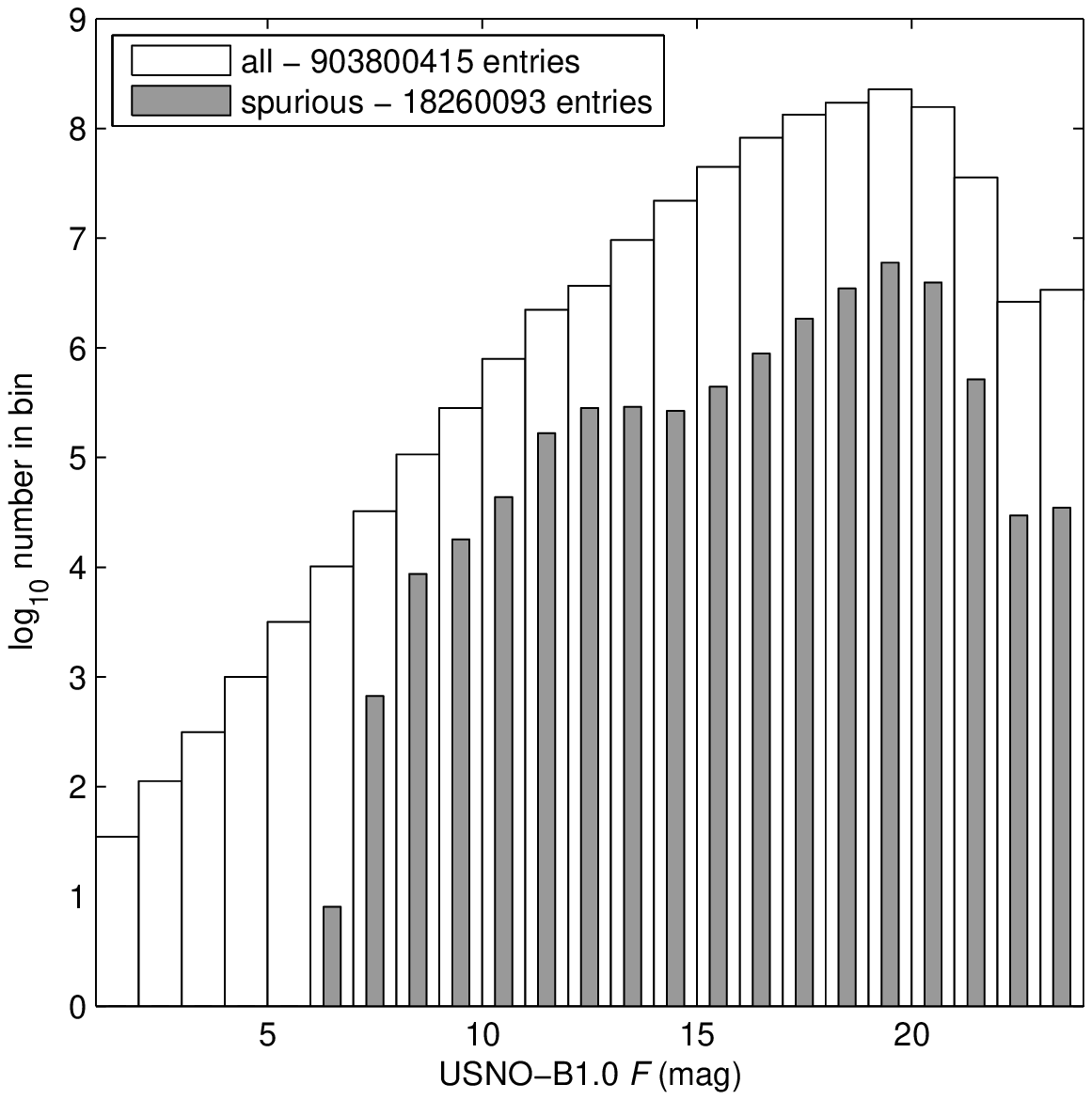}}\\
\resizebox{\twowidth}{!}{\includegraphics{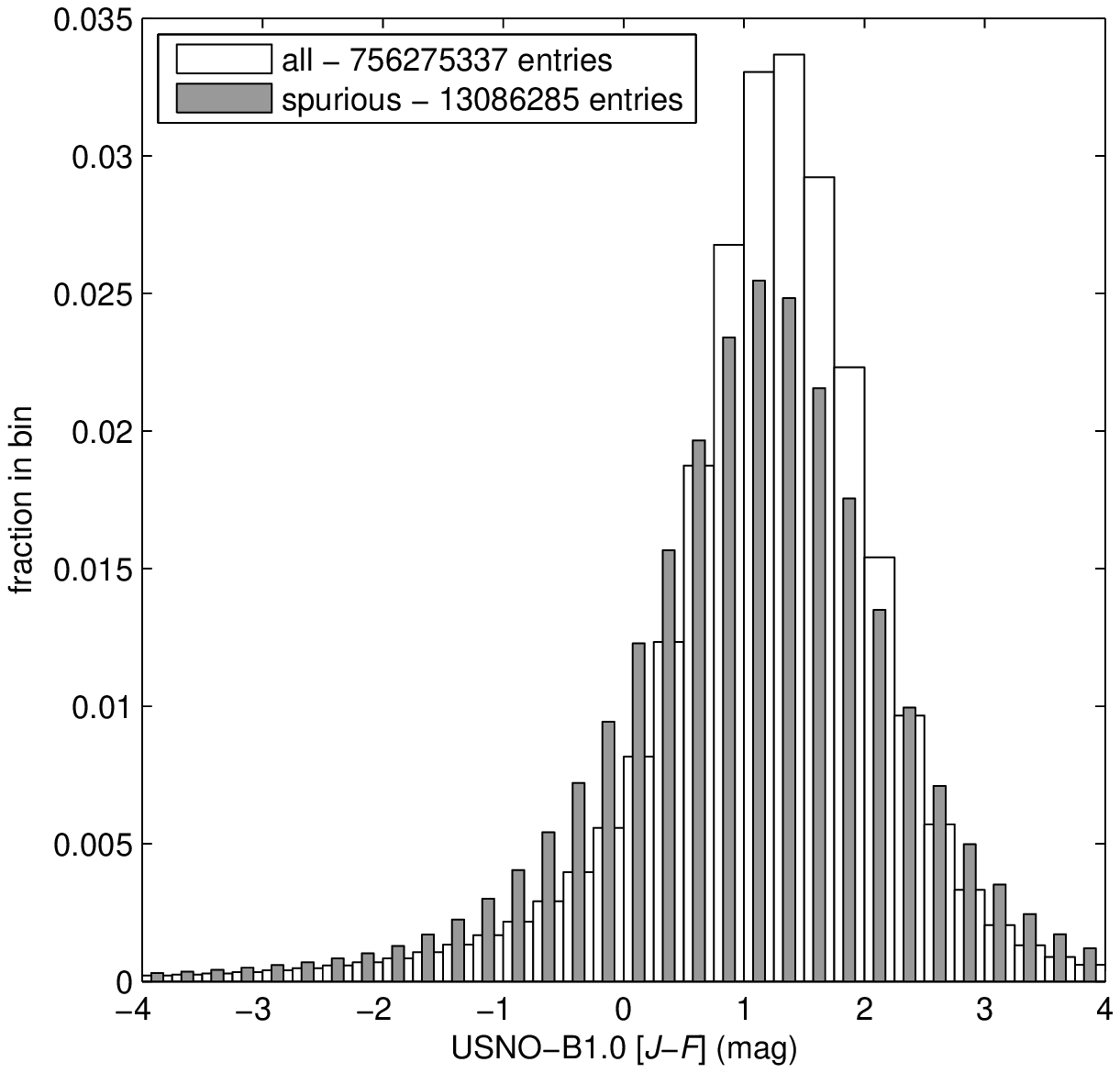}}%%
\resizebox{\twowidth}{!}{\includegraphics{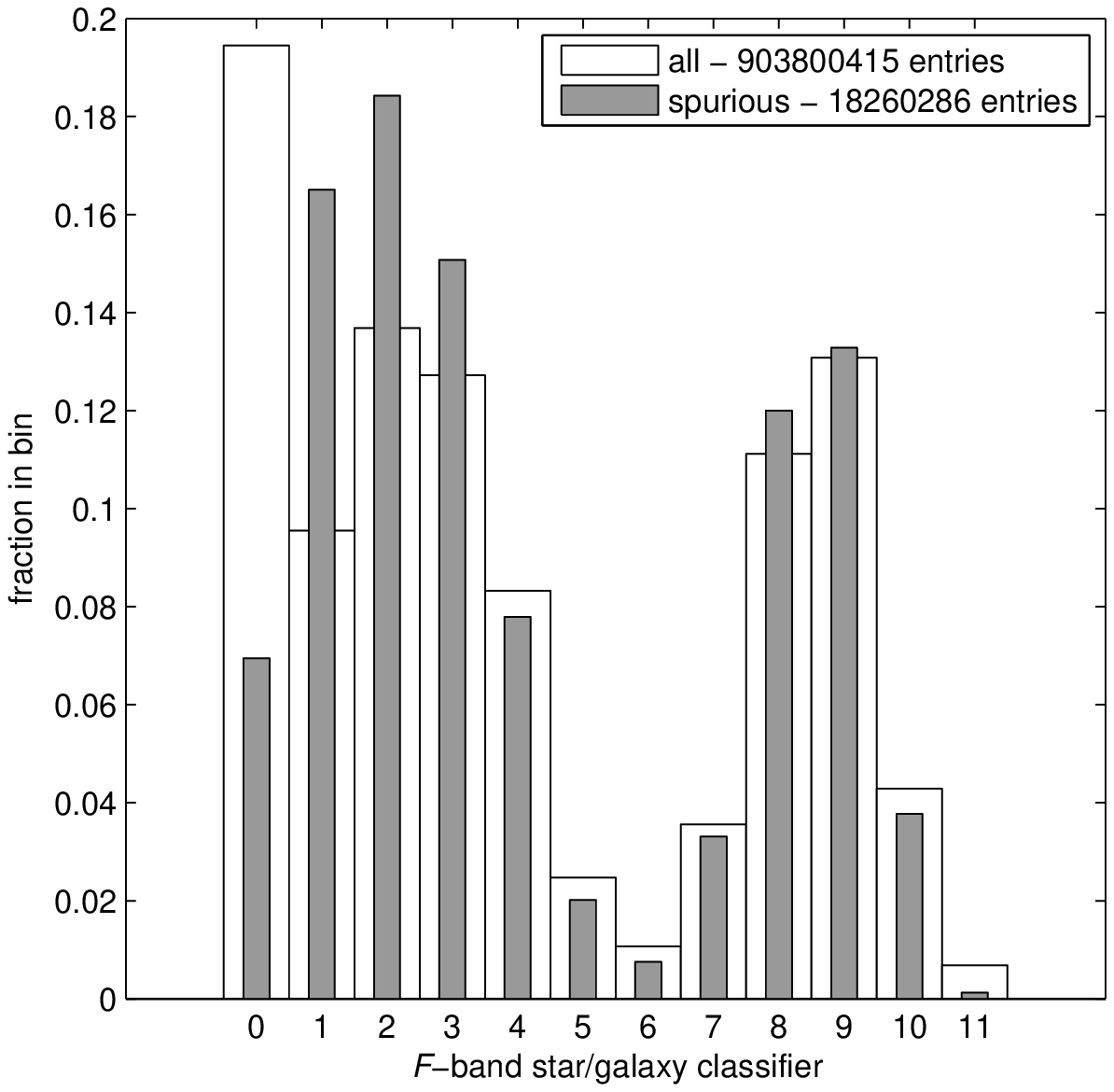}}
\caption{Spurious catalog entries are not obvious from their basic
photometric properties.
\textsl{Top-left panel:} Number of the five
bands in the \merged\ in which entries show detections, for all and
spurious entries.
\textsl{Top-right panel:} Magnitude distribution,
for all and spurious entries with detections in the $F$ band.
\textsl{Bottom-left panel:} J-F color distribution, for all and spurious
entries with detections in the $J$ and $F$ bands.
\textsl{Bottom-right panel:} \USNOB\ $F$-band star-galaxy separator
quantity, for all and spurious entries with detections in the $F$ band.%
\label{fig:spuriousProperties}}
\end{figure}

At the outset, we imagined that we could remove these spurious entries
trivially using the photometric properties listed in the Catalog.  For
example, there is no reason in principle that a spurious entry would
obtain a reasonable color or pass star--galaxy separation.  In
Figure~\ref{fig:spuriousProperties}, we show the distribution of the
spurious entries in photometric properties such as magnitude, color,
and star--galaxy separation.  This Figure shows that it would not have
been possible to identify the spurious on photometric grounds,
including even the \emph{number} of images with detections.
Presumably the reasonable colors and large numbers of overlapping
images in which the stars are detected result from the great stability
of the hardware and software employed in the construction of the
\USNOB.  It would have been extremely difficult to reliably identify
the spurious entries without automatic computer-vision techniques like
those employed in this project.

Associated with this paper is a small amount of computer code, the
information required to clean the \USNOB\ of the spurious entries we
identified, and some methods for accessing our cleaned version of the
\USNOB.  All of these are available at the \an\ web
site\footnote{http://astrometry.net/cleanusnob/}.

\acknowledgments We are very grateful to Dave Monet and the team that
created the \USNOB, which is one of astronomy's most productive and
useful resources.  We benefitted from useful discussions with Mike
Blanton, Keir Mierle, and David Warde-Farley, and from the
constructive comments of our anonymous referee.  DWH was partially
supported by the National Aeronautics and Space Administration (NASA;
grant NAG5-11669) and the National Science Foundation (NSF; grant
AST-0428465).  This research made use of the NASA Astrophysics Data
System, and the US Naval Observatory Precision Measuring Machine Data
Archive.

\clearpage

\clearpage

\clearpage

\clearpage

\clearpage

\end{document}